\documentstyle [floats,epsf,aps,eqsecnum,amstex]{revtex}
\draft
\global\firstfigfalse

\newcommand{\eqnref}[1]{Eq.~(\ref{#1})}
\newcommand{\eqnlessref}[1]{(\ref{#1})}

\def\Re{{\rm Re}}
\def\Im{{\rm Im}}
\def\Det{{\rm Det}}
\def\sech{{\rm sech}}
\def\Tr{{\rm Tr}}
\def\scrD{{\cal D}}
\def\scrK{{\cal K}}
\def\scrA{{\cal A}}
\def\ehat{{\bf \hat e}}

\begin{document}

\preprint{UW/PT-00-09}

\title{Effective String Theory of Vortices and Regge Trajectories}

\author{M. Baker and R. Steinke}

\address{%
	Department of Physics \\
	University of Washington \\
	Seattle, WA 98195-1560
	}%

\maketitle

\begin {abstract}
	{%
Starting from a field theory containing classical vortex
solutions, we obtain an effective string theory
of these vortices as a path integral over the
two transverse degrees of freedom of the string.
We carry out a semiclassical expansion of
this effective theory, and use it to obtain
corrections to Regge trajectories due to
string fluctuations.
	}%
\end {abstract}


\section{Introduction}

The goal of this paper is to derive an effective string theory of
vortices beginning with a field theory containing classical vortex
solutions. The Abelian Higgs model is an example of such a theory.
Nielsen and Olesen~\cite{Nielsen+Olesen} showed that this model has
classical magnetic vortex solutions. These vortices are tubes of
magnetic flux with constant energy per unit length.

The  motivation for this work came from the dual superconductor
picture of confinement~\cite{Nambu,Mandelstam,tHooft}.
In this picture, a dual Meissner effect confines
electric color flux ($Z_3$ flux) to narrow tubes connecting
quark--antiquark pairs. Calculations with explicit
models of this type~\cite{Baker2} have been compared
both with experimental data and with Monte Carlo simulations of
QCD~\cite{Bali}.
To a good approximation, aside from a color factor, the dual
Abelian Higgs model, coupling dual potentials to a scalar
Higgs field carrying magnetic charge, can be used to describe the
results of these calculations. However, these calculations
neglect the effect of fluctuations in the shape of the flux
tube on the $q \bar q$ interactions. We show in this paper
that taking account of those fluctuations leads to an effective
string theory of long distance QCD.

Well before the introduction of the idea of dual superconductivity,
string models~\cite{Isgur+Paton} had been used to
understand the origin of Regge trajectories, and they have
continued to be used to describe other features of hadron physics,
such as the spectrum of hybrid mesons. In the dual superconductor
picture, a string arises because the dual potentials couple to
a quark--antiquark pair via a Dirac string whose ends are a
source and sink of electric color flux. The effect of the string
is to create a flux tube (or Abrikosov--Nielsen--Olesen
vortex~\cite{Abrikosov,Nielsen+Olesen}) connecting the
quark--antiquark pair. As the pair moves, this flux tube
sweeps out a space time surface on which the dual Higgs field
must vanish. This
condition determines the location of the QCD string
in the dual superconductor picture.

The effort to obtain an effective string theory for
Abrikosov--Nielsen--Olesen vortices has a long history,
independent of any connection to QCD. Nambu~\cite{Nambu}
attached quarks to the ends of superconducting vortices,
and found an expression for the classical action
of the resulting ANO vortex in the singular London
limit of infinite Higgs mass.
He introduced a cutoff to render this action finite, and
showed that it was proportional to the area of the
worldsheet (the Nambu--Goto action).

F\"orster~\cite{Forster} took into consideration the curvature of
the worldsheet. He showed that in the strong coupling limit,
with the ratio of vector and scalar masses held fixed, the
effects of curvature were unimportant, and the classical action for
the vortex reduced to the Nambu--Goto action.
This limit can be regarded as the long distance limit, since only
zero mass excitations are left in the theory.
Equivalently, since the flux tube radius vanishes in
this limit, all physical distances, measured in units
of the flux tube radius, are becoming large. All
degrees of freedom except the transverse oscillations
of the vortex are frozen out.

Gervais and Sakita~\cite{Gervais+Sakita} first considered the
quantum theory of the vortices of the Abelian Higgs model
in the same long distance limit. They used the results of
F\"orster to define collective coordinates for the vortices,
by means of which they constructed an effective vortex action.
They also obtained a formal expression for the Feynman path
integral of the Abelian Higgs
model as an integration over vortex sheets.
However, they were not able to write this expression
as an integral over the physical degrees of freedom of
the vortices.

L\"uscher, Symanzik, and Weisz~\cite{Luscher1} considered the
leading semiclassical corrections to the classical Nambu--Goto action
due to transverse string fluctuations, and showed how to regulate
the resulting divergences. They showed that for a
string of length $R$ with fixed ends, the leading
semiclassical contribution to the heavy quark potential
is $-\pi/12R$. In a second paper, L\"uscher~\cite{Luscher2}
showed that this result was unaffected by the
addition of other terms to the effective string action.

Polchinski and Strominger~\cite{Pol+Strom} discussed
the relation of the Abelian Higgs model to fundamental
string theory, regarding the
theory of ANO vortices as an effective string theory.
They explained how existing string quantization
methods were inappropriate for quantizing the
vortices. To compensate for the anomalies~\cite{Polyakov} in these
quantization methods, they introduced an additional
term, the ``Polchinski Strominger term,'' into
the effective vortex action.

Akhmedov, Chernodub, Polikarpov, and Zubkov~\cite{ACPZ}
 studied the quantum theory of ANO vortices in the London limit,
In particular, they studied the transformation from field degrees of
freedom to vortex degrees of freedom. They showed that
the Jacobian of this transformation contained the
``Polchinski Strominger term'' as a factor. Although they,
like Gervais and Sakita, did not obtain a complete expression
for the path integral, this paper provided an important stimulus
to our own work.

In the current paper, we simplify and extend work done in
an earlier paper~\cite{Baker+Steinke}.
We begin with the path integral representation of
a field theory having vortex solutions. It is an effective
field theory describing phenomena at distances greater than the
flux tube radius. We end up with an
effective string theory of vortices
in a form suitable for explicit calculations.

We apply this theory to calculate the energy $E$
and angular momentum $J$ of the fluctuations of
a string bounded by the curve generated by the worldlines
of a quark--antiquark pair separated by a fixed distance and
rotating with fixed angular velocity. This gives the
contribution of string fluctuations to the Regge trajectory
$J(E^2)$, which we compare with the experimental
$\rho$ and $\omega$ trajectories.

\section{Outline}

In section \ref{partition section}, we rewrite the path integral
over field configurations of the Abelian Higgs model containing
vortices as an integral over surfaces on which the Higgs field vanishes.
This introduces a Jacobian due to the change from
field variables to string variables (surfaces). This Jacobian is
the key to determining the action of the effective string theory,
and to defining the integral over all surfaces.
We next use the formalism described in section~\ref{partition section}
to obtain an effective theory of ANO vortices.
In section \ref{Jacobian factor}, we show how
the Jacobian divides into a field part and a string
part. The two parts of the Jacobian
play different roles in the effective theory.
In section \ref{effective action section}, we define an
expression for the action of the effective string theory.
All the dependence on the Abelian Higgs model is contained
in the string action. We also obtain an expression
for the path integral over vortices.
In section \ref{coordinate fix}, we show how
to express the integral over surfaces as an integral over
the two physical degrees of freedom of the vortex, and obtain
the final form of the effective string theory.

In the remaining sections
we compute the leading semiclassical contribution to Regge
trajectories due to the fluctuations of the string.
We obtain an expression for the contribution of string
fluctuations to the effective action
in section \ref{small f section}, and in
sections \ref{regularization section}
and \ref{W_2 section} describe how to regularize this
expression, making use of the results of
L\"uscher, Symanzik, and Weisz~\cite{Luscher1}.
In section \ref{regge effective potential section} we
calculate the contribution of string fluctuations to the
effective action for a straight, rotating string, and
in section \ref{regge plot section} obtain the
resulting corrections to Regge trajectories.

\section{The Transformation from Fields to Strings}

\label{partition section}

In this section we consider the Abelian Higgs model
coupled via a Dirac string to a moving quark--antiquark
pair. We transform the path integral
over field configurations containing vortices to
an integral over the surfaces $\tilde x^\mu$
determining the location of the vortices.

\begin {figure}[ht]
	\epsfxsize=1.8in
	\epsfysize=1.8in
	\leavevmode{\hfill \epsfbox{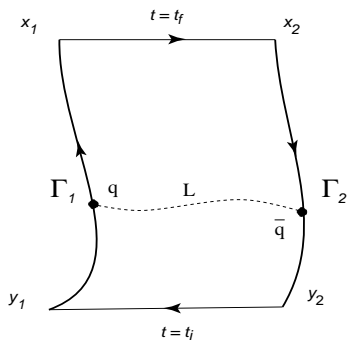} \hfill}
	\medskip
	\caption{The Loop $\Gamma$}
	\label{loop figure}
\end {figure}

We denote the (dual) potentials by $C_\mu$ and the
complex (monopole) Higgs field by $\phi$. The dual coupling
constant is $g = 2\pi/e$, where $e$ is the Yang--Mills
coupling constant ($\alpha_s = e^2/4\pi$).
The worldlines of the quark and antiquark
trajectories form the closed loop $\Gamma$ (see
Fig.~\ref{loop figure}).
The moving quark--antiquark pair
couples to the dual potentials $C_\mu$ via a Dirac string tensor
$G_{\mu\nu}^S$, which is nonvanishing along some line $L$ connecting
the $q \bar q$ pair. As the pair moves, the line $L$ sweeps out
a worldsheet $\tilde x^\mu(\xi)$ parameterized by coordinates
$\xi^a$, $a=1,2$. The field $\phi$ vanishes on this
worldsheet,
\begin{equation}
\phi(x^\mu) = 0 \,, \kern 0.3 in \hbox{ at } x^\mu = \tilde x^\mu(\xi) \,.
\end{equation}
The corresponding Dirac string tensor $G_{\mu\nu}^S$ is
given by
\begin{equation}
G_{\mu\nu}^S = - e \int d^2\xi \frac{1}{2} \epsilon^{ab}
\epsilon_{\mu\nu\alpha\beta} \frac{\partial\tilde x^\alpha}{\partial\xi^a}
\frac{\partial\tilde x^\beta}{\partial\xi^b}
\delta^{(4)}\left( x^\mu - \tilde x^\mu(\xi) \right) \,.
\label{G string def}
\end{equation}
The action $S$ of a field configuration which has a vortex
on the sheet $\tilde x^\mu(\xi)$ is
\begin{equation}
S = \frac{4}{3} \int d^4x \left[ -\frac{1}{4} \left(G_{\mu\nu}\right)^2
- \frac{1}{2} \left| (\partial_\mu - igC_\mu) \phi \right|^2
- \frac{\lambda}{4} \left( |\phi|^2 - \phi_0^2 \right)^2 \right] \,,
\label{firstaction}
\end{equation}
where the field strength $G_{\mu\nu}$ is given by
\begin{equation}
G_{\mu\nu} = \partial_\mu C_\nu - \partial_\nu C_\mu + G_{\mu\nu}^S \,.
\label{G mu nu def}
\end{equation}
The Higgs mechanism gives the vector particle (dual gluon) a mass
$M_V = g\phi_0$
and the scalar particle a mass $M_S = \sqrt{2\lambda}\phi_0$, where
$\phi_0$ is the vacuum expectation value of the Higgs field.
We have introduced the color factor $\frac{4}{3}$ in
\eqnlessref{G mu nu def} because we are interested in using
$S$ as a model for long distance QCD.
We consider $S$ to be an effective action describing
distances greater than the flux tube radius $a$.

The long distance $q \bar q$ interaction is determined by
the Wilson loop $W[\Gamma]$,
\begin{equation}
W[\Gamma] = \int \scrD\phi^* \scrD\phi \scrD C^{\mu}
e^{i(S[\phi,C] + S_{GF})} \,,
\label{originalpartition}
\end{equation}
where $S_{GF}$ is a gauge fixing term. The functional integrals
are cut off at the momentum scale $1/a$.
The action \eqnlessref{firstaction} describes a field theory
having classical vortex solutions. The functional
integral \eqnlessref{originalpartition} goes over
all field configurations containing a vortex
bounded by $\Gamma$.

Previous calculations~\cite{Baker2} of $W[\Gamma]$ were carried out
in the classical
approximation (corresponding to a flat vortex sheet $\tilde x^\mu$),
and showed that the Landau--Ginzburg parameter
$\lambda/g^2$ is approximately equal to $\frac{1}{2}$. This corresponds
to a superconductor on the border between type I and type II. In
this situation, both particles have the same mass $M = M_V = M_S$,
the string tension is $\sigma = \frac{4}{3} \pi \phi_0^2$,
and the flux tube radius is $a=\sqrt{2}/M$.

To take into account the fluctuations of these vortices, we must
evaluate $W[\Gamma]$ beyond the classical approximation.
We carry out the functional
integration \eqnlessref{originalpartition} in two steps:
(1) We fix the location of a vortex sheet $\tilde x^\mu$,
and integrate only over field configurations for which $\phi(x^\mu)$
vanishes on $\tilde x^\mu$. (2) We integrate
over all possible vortex sheets.
To implement this procedure, we introduce into the
functional integral \eqnlessref{originalpartition}
the factor one, written in the form
\begin{equation}
1 = J[\phi] \int \scrD \tilde x^\mu
\delta\left[\Re\phi(\tilde x^{\mu}(\xi))\right]
\delta\left[\Im\phi(\tilde x^{\mu}(\xi))\right] \,.
\label{inserttildex}
\end{equation}
The integration $\scrD \tilde x^\mu$ is over the four
functions $\tilde x^\mu(\xi)$. The functions $\tilde x^\mu(\xi)$
are a particular parameterization of the worldsheet $\tilde x^\mu$.

The expression \eqnlessref{inserttildex} implies that the string worldsheet
$\tilde x^\mu$, determined by the $\delta$ functions, is the
surface of the zeros of the field $\phi$. The factor $J[\phi]$ is a Jacobian,
and is defined by \eqnref{inserttildex}. Inserting \eqnlessref{inserttildex}
into \eqnlessref{originalpartition} puts the Wilson loop in the form
\begin{equation}
W[\Gamma] = \int \scrD\phi^* \scrD\phi \scrD C^{\mu} e^{i(S[\phi,C]
+ S_{GF})} J[\phi] \int \scrD \tilde x^\mu
\delta\left[\Re\phi(\tilde x^{\mu}(\xi))\right]
\delta\left[\Im\phi(\tilde x^{\mu}(\xi))\right] \,.
\label{beforeswitch}
\end{equation}
We then reverse the order of the field integration and the string
integration over surfaces $\tilde x^\mu(\xi)$,
\begin{equation}
W[\Gamma] = \int \scrD \tilde x^\mu \int \scrD\phi^* \scrD\phi
\scrD C^{\mu} J[\phi] \delta\left[\Re\phi(\tilde x^{\mu}(\xi))\right]
\delta\left[\Im\phi(\tilde x^{\mu}(\xi))\right] e^{i(S[\phi,C]
+ S_{GF})} \,.
\label{afterswitch}
\end{equation}
In \eqnref{beforeswitch}, the $\delta$ functions fix $\tilde x^\mu$
to lie on the surface of the zeros of a given field $\phi$, while in
\eqnref{afterswitch}, they restrict the field $\phi$ to vanish on a
given surface $\tilde x^\mu$. The integral over $\phi$ in
\eqnref{afterswitch} is therefore restricted to functions $\phi$ which
vanish on $\tilde x^\mu$, in contrast to the integral over $\phi$ in
\eqnref{beforeswitch}, in which $\phi$ can be any function.

\section{Factorization of the Jacobian}

\label{Jacobian factor}

To evaluate $W[\Gamma]$ we divide $J[\phi]$ into two parts.
The Jacobian $J[\phi]$ in \eqnref{afterswitch} is evaluated
for field configurations $\phi$ which vanish on a
particular surface $\tilde x^\mu$.
We make this explicit by writing \eqnlessref{inserttildex} as
\begin{equation}
J[\phi,\tilde x^{\mu}]^{-1} = \int \scrD \tilde y^\mu
\delta\left[\Re\phi(\tilde y^{\mu}(\tau))\right]
\delta\left[\Im\phi(\tilde y^{\mu}(\tau))\right] \,,
\label{Jdef}
\end{equation}
where $\tilde y^\mu$ is some other string worldsheet, distinct from
$\tilde x^\mu$. The evaluation of the Jacobian is the essential new
ingredient in deriving $W[\Gamma]$.

The $\delta$ functions in \eqnlessref{Jdef} select
surfaces $\tilde y^\mu(\tau)$ which lie in a neighborhood of
the surface $\tilde x^\mu(\xi)$ of the zeros of $\phi$.
We separate
$\tilde y^\mu(\tau)$ into components lying on the
surface $\tilde x^\mu(\xi)$ and components lying along
vectors $n_\mu^A(\xi)$ normal to $\tilde x^\mu(\xi)$ at the point $\xi$:
\begin{equation}
\tilde y^\mu(\tau) = \tilde x^\mu(\xi(\tau)) + y_\perp^A(\xi(\tau))
n_\mu^A(\xi(\tau)) \,.
\label{localcoords}
\end{equation}
\begin{figure}[ht]
	\epsfxsize=3in
	\leavevmode{\hfill \epsfbox{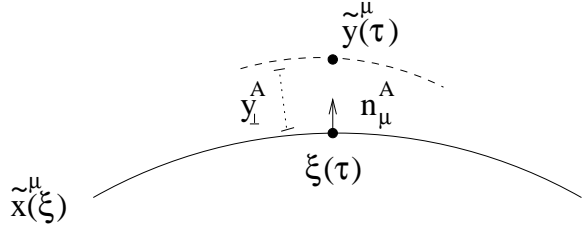} \hfill}
	\medskip
	\caption{Worldsheets and normal vectors}
	\label{normal figure}
\end{figure}
The point $\tilde x^\mu(\xi(\tau))$ is the point on the
surface $\tilde x^\mu(\xi)$ lying closest to $\tilde y^\mu(\tau)$,
and the magnitude of $y_\perp^A(\xi(\tau))$ is the distance
from $\tilde y^\mu(\tau)$ to $\tilde x^\mu(\xi(\tau))$
(see Fig.~\ref{normal figure}).

We evaluate the Jacobian \eqnlessref{Jdef} by making the change
of variables
\begin{equation}
\tilde y^\mu(\tau) \to \left( \xi(\tau) , y_\perp^A(\xi) \right)
\label{transformation}
\end{equation}
defined by \eqnlessref{localcoords}.
Although the $\delta$ functions in \eqnlessref{Jdef} force $y_\perp^A$
to vanish, the integrations over $y_\perp^A$ give a contribution to
the Jacobian.
Furthermore, this contribution depends on the field variable $\phi$ in
a neighborhood of the surface. The integration over the
reparameterizations $\xi(\tau)$ of the surface $\tilde x^\mu(\xi)$,
on the other hand, depends upon the surface, but not on the fields.
The change of variables \eqnlessref{transformation} leads to a
factorization of the Jacobian into a field contribution, and into
a contribution
depending only on the intrinsic properties of the worldsheet $\tilde
x^\mu(\xi)$.

We now exhibit the factorization of the Jacobian.
Under the transformation \eqnlessref{transformation}, the integral
over $\tilde y^\mu$ becomes
\begin{eqnarray}
\scrD \tilde y^\mu &=& \Det_\tau\left[\epsilon^{\mu\nu\alpha\beta}
\frac{1}{2} \epsilon^{ab} \frac{\partial \tilde x^\mu}{\partial\xi^a}
\frac{\partial \tilde x^\nu}{\partial\xi^b}
\frac{1}{2} \epsilon^{AB} n_\alpha^A n_\beta^B\right]
\scrD y_\perp^A \scrD\xi
\nonumber \\
&=& \Det_\tau\left[\sqrt{ -\frac{1}{2} \left( \epsilon^{ab}
\frac{\partial \tilde x^\mu}{\partial\xi^a}
\frac{\partial \tilde x^\nu}{\partial\xi^b} \right)^2
\frac{1}{2} \left( \epsilon^{AB} n_\alpha^A n_\beta^B \right)^2}\right]
\scrD y_\perp^A \scrD\xi
\nonumber \\
&=& \Det_\tau\left[\sqrt{-g(\xi)}\Big|_{\xi=\xi(\tau)}\right]
\scrD y_\perp^A \scrD\xi \,,
\label{y Jacobian}
\end{eqnarray}
where $\sqrt{-g}$ is the square root of the determinant
of the induced metric
\begin{equation}
g_{ab} = \frac{\partial \tilde x^\mu}{\partial\xi^a}
\frac{\partial \tilde x^\mu}{\partial\xi^b}
\end{equation}
evaluated on the worldsheet $\tilde x^\mu$.
Appendix~\ref{notation appendix} gives
a summary of our notation, and of the relations used to obtain
\eqnlessref{y Jacobian}.

The functional
determinant in \eqnlessref{y Jacobian} is the product of its
argument evaluated at all points $\tau$ on the sheet,
in the same way that the integration over $\scrD\tilde y^\mu$ is a product
of integrals at all points $\tau$.
Making the change of coordinates \eqnlessref{localcoords},
\eqnlessref{transformation} in the Jacobian \eqnlessref{Jdef} gives
\begin{eqnarray}
J[\phi,\tilde x^{\mu}]^{-1}
&=& \int \scrD \xi \scrD y_{\perp}^A
\Det_\tau[\sqrt{-g}]
\delta\left[\Re\phi\left(\tilde x^{\mu}(\xi(\tau)) +
y_{\perp}^A(\xi(\tau)) n_A^{\mu}(\xi(\tau))\right)\right]
\nonumber \\
& & \times \delta\left[\Im\phi\left(\tilde x^{\mu}(\xi(\tau))
+ y_{\perp}^A(\xi(\tau)) n_A^{\mu}(\xi(\tau))\right)\right] \,.
\label{J partial split}
\end{eqnarray}

\eqnref{J partial split} has the form:
\begin{equation}
J[\phi,\tilde x]^{-1} = \int \scrD \xi(\tau)
\Det_{\tau}\left[\sqrt{-g}\right] J_{\perp}[\phi,
\tilde x^{\mu}(\xi(\tau))]^{-1} \,,
\label{newJacob}
\end{equation}
where
\begin{eqnarray}
J_{\perp}[\phi,\tilde x^{\mu}(\xi(\tau))]^{-1} &=& \int \scrD y_{\perp}^A
\delta\left[\Re\phi\left(\tilde x^{\mu}(\xi(\tau)) + y_{\perp}^A(\xi(\tau))
n_A^{\mu}(\xi(\tau))\right)\right]
\nonumber \\
& & \times \delta\left[\Im\phi\left(\tilde x^{\mu}(\xi
(\tau)) + y_{\perp}^A(\xi(\tau)) n_A^{\mu}(\xi(\tau))\right)\right]
\end{eqnarray}
contains all the dependence on $\phi$. Since $J_\perp$ is
independent of the parameterization $\xi(\tau)$, the
Jacobian factors into two parts:
\begin{equation}
J[\phi,\tilde x]^{-1} = J_{\parallel}[\tilde x]^{-1} J_{\perp}[\phi,
\tilde x]^{-1} \,,
\label{Jfactor}
\end{equation}
where
\begin{equation}
J_{\parallel}[\tilde x]^{-1} = \int \scrD \xi
\Det_{\tau}\left[\sqrt{-g}\right] \,.
\label{Jparallel}
\end{equation}
The string part $J_\parallel$
of the Jacobian arises from the parameterization
degrees of freedom. In the next section, we show
that $J_\perp$ is the Faddeev--Popov determinant for the
$\delta$ functions in \eqnlessref{afterswitch}.
This allows us to define the action of the
effective string theory.
In the following section, we will use $J_\parallel$ to fix
the reparameterization degrees of freedom.

\section{The String Action}

\label{effective action section}

Inserting the factorized form \eqnlessref{Jfactor} of $J[\phi]$
into the expression \eqnlessref{afterswitch} for $W[\Gamma]$
gives the Wilson Loop the form
\begin{equation}
W[\Gamma] = \int \scrD \tilde x^\mu J_\parallel[\tilde x]
e^{iS_{{\rm eff}}} \,,
\label{stringrep}
\end{equation}
where the action $S_{{\rm eff}}$ of the effective string theory is
given by
\begin{equation}
e^{iS_{{\rm eff}}[\tilde x^\mu(\xi)]} = \int \scrD\phi^* \scrD\phi
\scrD C^\mu J_\perp[\phi] \delta\left[\Re\phi(\tilde x^\mu(\xi))\right]
\delta\left[\Im\phi(\tilde x^\mu(\xi))\right] e^{i(S + S_{GF})} \,.
\label{Seff}
\end{equation}
The string action \eqnlessref{Seff} was obtained previously by Gervais
and Sakita~\cite{Gervais+Sakita}. The novel feature of our
result is the string integration measure of the Wilson loop
\eqnlessref{stringrep}.

The string action depends upon the field part $J_\perp$
of the Jacobian,
\begin{equation}
J_{\perp}[\phi,\tilde x^{\mu}]^{-1} = \int \scrD y_{\perp}^A
\delta\left[\Re\phi\left(\tilde x^{\mu} +
y_{\perp}^A n_{\mu A} \right)\right]
\delta\left[\Im\phi\left(\tilde x^{\mu} +
y_{\perp}^A n_{\mu A} \right)\right] \,.
\label{J perp of y}
\end{equation}
The $\delta$ functions force $y_{\perp}^A$ to be zero, so we can expand
their arguments in a power series in $y_{\perp}^A$,
\begin{equation}
\phi\left(y^{\mu}\right) = \phi\left(\tilde x^{\mu}\right)
+ y_{\perp}^A n_A^{\nu} \partial_{\nu} \phi\left(\tilde x^{\mu}\right)
+ O\left( y_{\perp}^2 \right) \,.
\label{phi expand in y}
\end{equation}
The zeroth order term in \eqnlessref{phi expand in y}
vanishes because $\tilde x^{\mu}$ is the surface of the
zeros of $\phi$. The integration \eqnlessref{J perp of y}
over $y_\perp^A$ gives the result
\begin{equation}
J_{\perp}[\phi,\tilde x^{\mu}]^{-1} = \Det^{-1}_{\xi}\left[
\epsilon^{AB} n_A^{\mu} n_B^{\nu} \left(\partial_{\mu}\Re\phi\right)
\left(\partial_{\nu}\Im\phi\right) \Big|_{x^{\mu} = \tilde x^{\mu}} \right] \,.
\label{J_perp expressed}
\end{equation}
The Jacobian $J_\perp$ is a Faddeev--Popov determinant,
which we discuss in Appendix~\ref{J_perp FP appendix}.

\eqnref{Seff} gives the action $S_{{\rm eff}}(\tilde x^\mu)$
of the effective string theory as an integral over field configurations
which have a vortex fixed at $\tilde x^\mu$. Since the vortex
theory \eqnlessref{originalpartition} is an effective long distance
theory, the path integral \eqnlessref{originalpartition} for $W[\Gamma]$,
written in terms of the fields of the Abelian Higgs model,
is cut off at a scale $\Lambda$ which is on the order of
the mass $M$ of the dual gluon. Furthermore, the
integration \eqnlessref{stringrep} over $\tilde x^\mu$
includes all the long distance fluctuations of the theory.
Therefore, the path integral \eqnlessref{Seff}
contains neither short distance nor long distance fluctuations,
and is determined by minimizing the field action
$S[\tilde x^\mu, \phi, C_\mu]$ for a fixed position of the vortex sheet:
\begin{equation}
S_{{\rm eff}}[\tilde x^\mu] = S[\tilde x^\mu, \phi^{{\rm class}},
C_\mu^{{\rm class}}] \,, \kern 0.5 in \phi^{{\rm class}}(\tilde x^\mu) = 0 \,.
\label{classical string}
\end{equation}
The fields $\phi^{{\rm class}}$ and $C_\mu^{{\rm class}}$
are the solutions of the classical equations of motion,
subject to the boundary condition $\phi(\tilde x^\mu) = 0$.

The action $S_{{\rm eff}}$ depends both on the distance $R$
between the quarks, and the radius of curvature $R_V$ of the
vortex sheet bounded by $\Gamma$. In the long distance
limit, when the length of the string $R$ and its radius $R_V$
are large compared to the thickness of the flux tube $a$, the
string action \eqnlessref{classical string} becomes the
Nambu--Goto action $S_{NG}$,
\begin{equation}
S_{NG} = -\sigma \int d^2\xi \sqrt{-g} \,,
\label{NG action def}
\end{equation}
where $\sigma$ is the classical string tension,
determined from the solution of the Nielsen--Olesen equations
for a straight, infinitely long string.

It is convenient to separate the action \eqnlessref{classical string}
into its perturbative and nonperturbative parts:
\begin{equation}
S_{{\rm eff}}[\tilde x^\mu]
= S[\tilde x^\mu, \phi^{{\rm class}}, C_\mu^{{\rm class}}]
= S^{{\rm Maxwell}}[\tilde x^\mu] + S^{{\rm NP}}[\tilde x^\mu] \,,
\label{break Abelian action}
\end{equation}
where $S^{{\rm Maxwell}}$ is the action obtained by setting
$\lambda = g = 0$ in \eqnref{firstaction}. The value
of $S^{{\rm Maxwell}}$ depends only upon the boundary
$\Gamma$, and is the usual electromagnetic interaction
between charged particles:
\begin{equation}
S^{{\rm Maxwell}}[\Gamma] = \frac{4}{3} \frac{e^2}{2}
\oint dx^\mu \oint d{x'}^\mu {\cal D}_{\mu\nu}(x^\mu - {x'}^\mu) \,,
\label{maxwell action}
\end{equation}
where ${\cal D}_{\mu\nu}$ is the photon propagator.

To calculate the Wilson loop $W[\Gamma]$ from the
effective string theory \eqnlessref{stringrep},
we must also examine $S_{{\rm eff}}$ at
smaller values of $R$ and $R_V$, on the order
of the string thickness $a$. We first consider
the dependence of $S_{{\rm eff}}$ on $R$ for a flat string,
where $R_V \to\infty$. In this case, the curve $\Gamma$
is a rectangle of length $T$ in the time direction, and width
$R$ in the space direction. In the large $T$
limit, the action $S_{{\rm eff}}$
reduces to the product of $T$ and the potential
$V^{{\rm class}}(R)$ previously used to fit the energy levels of
heavy quark systems.

Evaluation of \eqnlessref{break Abelian action} for a
flat sheet gives a corresponding
decomposition of $V^{{\rm class}}(R)$,
\begin{equation}
V^{{\rm class}}(R) = V^{{\rm Coulomb}}(R) + V^{{\rm NP}}(R) \,.
\end{equation}
For small $R$,
\begin{equation}
V^{{\rm class}}(R) @>>{R\to 0}> V^{{\rm Coulomb}}(R) = -\frac{4}{3}
\frac{\alpha_s}{R} \,,
\end{equation}
while \eqnref{NG action def} gives the large $R$ behavior
\begin{equation}
V^{{\rm NP}}(R) @>>{R\to\infty}> \sigma R \,.
\label{V class long}
\end{equation}

Recent numerical studies~\cite{Gubarev+Polikarpov+Zakharov}
of the classical equations of motion for a flat
sheet have shown that for a superconductor on the I--II border, the
long distance behavior \eqnlessref{V class long} of
$V^{{\rm NP}}(R)$ persists to small values of $R$, even to
values less than the string thickness $a$. Therefore, for a
superconductor on the I--II border, $V^{{\rm class}}(R)$
is, to a good approximation, equal to the Cornell
potential~\cite{Eichten+Gottfried}:
\begin{equation}
V^{{\rm class}}(R) \approx -\frac{4}{3} \frac{\alpha_s}{R} + \sigma R \,.
\label{V class}
\end{equation}
In other words, for a flat sheet,
\begin{equation}
S^{{\rm NP}}(R) \approx S_{NG} \,.
\end{equation}
Thus, for short straight strings the Nambu--Goto action remains
a good approximation to the nonperturbative part of the classical
action for a superconductor on the type I--II border.

Next, consider the nonperturbative contribution to the
classical action for a long bent string. (The Maxwell action
has the value \eqnlessref{maxwell action} independent of the
shape of the vortex.) The leading correction to the
Nambu--Goto action when the string is bent is the curvature term:
\begin{equation}
S_{{\rm curvature}} = - \beta \int d^2\xi \sqrt{-g}
\left(\scrK^A_{ab}\right)^2 \,,
\label{curvature term}
\end{equation}
where $\scrK^A_{ab}$ is the extrinsic curvature.
\begin{equation}
\scrK^A_{ab} = n_\mu^A \partial_a \partial_b x^\mu \,.
\end{equation}
The magnitude of $\scrK^A_{ab}$ is of the order of $1/R_V$,
so that $S_{{\rm curvature}} \sim (a^2/R_V^2) S_{NG}$.

The calculation of the ``rigidity'' $\beta$ determining
the size of $S_{{\rm curvature}}$ has been considered by
a number of authors~\cite{Maeda}, but the value of $\beta$
for a superconductor on the I--II border was never calculated.
We conjecture that the value of $\beta$ is small, because
de Vega and Schaposnik~\cite{deVega+Schaposnik}
have shown that the components of the stress tensor
perpendicular to the axis of a straight Nielsen--Olesen
flux tube vanish for a superconductor on the border
between type I and type II. In other words, there are no ``bonds''
perpendicular to the field lines of a straight flux
tube of infinite extent. When the flux tube is bent slightly,
there are no perpendicular bonds to be stretched or
compressed, and the change in the energy is just the
string tension multiplied by the change in length.
That is, the curvature term, which in
a sense represents the attraction or
repulsion between neighboring parts of the
string, should vanish.
A more formal argument can be made by regarding the
borderline superconductor as the long distance
limit of a theory where the forces between vortices
become weak. Polyakov~\cite{morePolyakov} has
shown, using renormalization group methods, that
$\beta$ also vanishes in this limit.

Similar heuristic arguments give a reason for the above mentioned result
that the Nambu--Goto action is a good approximation
for short, straight strings on the I--II border.
The bending of the field lines as the quark--antiquark separation
becomes smaller causes no additional changes in the energy.

We therefore take the action of the effective string theory
to be equal to the sum of the Maxwell action~\eqnlessref{maxwell action}
and the Nambu--Goto action ~\eqnlessref{NG action def}:
\begin{equation}
S_{{\rm eff}}[\tilde x^\mu] = S^{{\rm Maxwell}}[\Gamma]
- \sigma \int d^2\xi \sqrt{-g} \,.
\label{given action}
\end{equation}
We use \eqnref{given action} for the full range of string
lengths $R$ and radii of curvature $R_V$ greater than the
inverse of the mass $M$ of the dual gluon, which is
the cutoff of the effective string theory~\eqnlessref{stringrep}.

\eqnref{given action} for $S_{{\rm eff}}[\tilde x^\mu]$ is
the generalization of \eqnlessref{V class}
to a general sheet. The first term,
$S^{{\rm Maxwell}}[\Gamma]$, is just a boundary term, independent
of the fluctuating string, and we take take
$S_{{\rm eff}} = S_{NG}$ for the calculations carried
out in the rest of this paper. In the next section we
show how to carry out the integration over $\tilde x^\mu$
in \eqnlessref{stringrep} by separating the degrees of freedom
of the worldsheet $\tilde x^\mu$ into two physical
degrees of freedom and two reparameterization degrees
of freedom. This treatment makes no use of \eqnlessref{given action},
and is applicable to any effective string theory of vortices.

\section{Effective Theory of Transverse String Fluctuations}

\label{coordinate fix}

We next show how to evaluate the integral over
$\tilde x^\mu(\xi)$ in \eqnref{stringrep},
\begin{equation}
W[\Gamma] = \int \scrD \tilde x^\mu J_\parallel[\tilde x^\mu]
e^{iS_{{\rm eff}}} \,.
\label{vortex partition}
\end{equation}
The integration over $\tilde x^\mu(\xi)$ is the product of an integral
over string worldsheets and an integral over reparameterizations
of the coordinates of the string. The Jacobian $J_\parallel$
is the inverse of the integration \eqnlessref{Jparallel} over
reparameterization degrees of freedom. In this section,
we fix the parameterization of the string, and show that
$J_\parallel$ cancels the integration over reparameterizations.

Any surface $\tilde x^\mu$ has only two physical degrees of
freedom. The other two degrees of freedom represent the invariance of
the surface under coordinate reparameterizations.
We fix the coordinate reparameterization symmetry by choosing
a particular ``representation'' $x_p^\mu$ of the surface, which
depends on two functions $f^1(\xi)$, $f^2(\xi)$,
\begin{equation}
x_p^\mu(\xi) = x_p^\mu[f^1(\xi), f^2(\xi), \xi] \,.
\label{x_p def}
\end{equation}
A particular example of a representation $x_p^\mu$ is obtained
by expanding in transverse fluctuations $x_\perp^A$ about a
fixed sheet $\bar x_p^\mu$,
\begin{equation}
x_p^\mu(\xi) = x_p^\mu[x_\perp^A(\xi),\xi] =
\bar x_p^\mu(\xi) + x_\perp^A(\xi) \bar n^\mu_A(\xi) \,.
\label{normal rep}
\end{equation}
The vectors $\bar n^\mu_A$ are orthogonal to the
surface $\bar x_p^\mu$.
In this example, $f^1$ and $f^2$ are the transverse coordinates
$x_\perp^A$.

Any physical surface can be expressed in terms of $x_p^\mu$
by a suitable choice of $f^1$ and $f^2$. In particular, the
worldsheet $\tilde x^\mu(\xi)$
appearing in the integral \eqnlessref{vortex partition}
can be written in terms of a reparameterization $\tilde\xi(\xi)$
of the representation $x_p^\mu$,
\begin{equation}
\tilde x^\mu(\xi) = x_p^\mu[f^1(\tilde\xi(\xi)),
f^2(\tilde\xi(\xi)), \tilde\xi(\xi)] \,.
\label{x tilde of x_p}
\end{equation}
The four degrees of freedom in $\tilde x^\mu(\xi)$ are
replaced by two physical degrees of freedom $f^1(\xi)$, $f^2(\xi)$
and two reparameterization degrees of freedom $\tilde\xi(\xi)$.

We can write the integral over $\tilde x^\mu(\xi)$
in \eqnlessref{vortex partition} in terms of integrals over
$f^1(\xi),f^2(\xi)$ and $\tilde\xi(\xi)$,
\begin{equation}
\scrD \tilde x^\mu = \Det\left[\epsilon_{\mu\nu\alpha\beta}
\frac{1}{2} \epsilon^{ab} \frac{\partial\tilde x^\mu}{\partial\tilde\xi^a}
\frac{\partial\tilde x^\nu}{\partial\tilde\xi^b}
\frac{\partial\tilde x^\alpha}{\partial f^1}
\frac{\partial\tilde x^\beta}{\partial f^2}\right]
\scrD f^1 \scrD f^2 \scrD\tilde\xi \,.
\end{equation}
Noting that the
derivative of $\tilde x^\mu$ with respect to $\tilde\xi(\xi)$ is,
\begin{equation}
\frac{\partial\tilde x^\mu}{\partial\tilde\xi^a}
= \left[ \frac{\partial x_p^\mu}{\partial f^1} \frac{\partial f^1}
{\partial\xi^a} + \frac{\partial x_p^\mu}{\partial f^2} \frac{\partial f^2}
{\partial\xi^a} + \frac{\partial x_p^\mu}{\partial\xi^a} \right]
\Bigg|_{\xi=\tilde\xi} \,,
\end{equation}
we can write
\begin{eqnarray}
\scrD \tilde x^\mu &=& \Det\left[ \epsilon_{\mu\nu\alpha\beta} \frac{1}{2}
\epsilon^{ab} \frac{\partial x_p^\alpha}{\partial\xi^a}
\frac{\partial x_p^\beta}{\partial\xi^b}
\frac{\partial x_p^\mu}{\partial f^1} \frac{\partial x_p^\nu}
{\partial f^2} \Bigg|_{\xi=\tilde\xi} \right]
\scrD f^1 \scrD f^2 \scrD \tilde\xi
\nonumber \\
&=& \Det\left[ \tilde t_{\mu\nu} \sqrt{-g^p}
\frac{\partial x_p^\mu}{\partial f^1} \frac{\partial x_p^\nu}
{\partial f^2} \Bigg|_{\xi = \tilde\xi} \right]
\scrD f^1 \scrD f^2 \scrD \tilde\xi \,,
\label{measure shift}
\end{eqnarray}
where
\begin{equation}
\tilde t_{\mu\nu} = \frac{1}{2} \epsilon_{\mu\nu\alpha\beta}
\frac{\epsilon^{ab}}{\sqrt{-g}} \frac{\partial\tilde x^\alpha}{\partial\xi^a}
\frac{\partial\tilde x^\beta}{\partial\xi^b} \,,
\end{equation}
is the antisymmetric tensor normal to the worldsheet and,
\begin{equation}
g_{ab}^p = \frac{\partial x_p^\mu}{\partial\xi^a}
\frac{\partial x_p^\mu}{\partial\xi^b} \,,
\end{equation}
is the induced metric of $x_p^\mu(\xi)$. The metric $g_{ab}^p$
is related to the metric $g_{ab}$ of $\tilde x^\mu(\xi)$,
\begin{equation}
g_{ab} = \frac{\partial\tilde\xi^c}{\partial\xi^a}
\frac{\partial\tilde\xi^d}{\partial\xi^b} g_{cd}^p \,.
\end{equation}
The induced metric $g_{ab}$ of the original worldsheet $\tilde x^\mu(\xi)$
does not appear in \eqnref{measure shift}
because the determinant is independent of $\tilde\xi$.
Only the induced metric $g_{ab}^p$ of the worldsheet
$x_p^\mu(\xi)$ enters into the determinant.

With the parameterization \eqnlessref{x tilde of x_p}
of $\tilde x^\mu$,
the path integral \eqnlessref{vortex partition} takes the form
\begin{equation}
W[\Gamma] = \int \scrD\tilde\xi \scrD f^1 \scrD f^2
\Det\left[ \tilde t^{\mu\nu} \frac{\partial x_p^\mu}
{\partial f^1} \frac{\partial x_p^\nu} {\partial f^2} \right]
\Det[\sqrt{-g^p}] J_\parallel e^{iS_{{\rm eff}}} \,.
\label{W subbed}
\end{equation}
The action $S_{{\rm eff}}$ is parameterization independent,
so it is independent of $\tilde\xi(\xi)$. The same is true
for $J_\parallel$. Furthermore, $\tilde t_{\mu\nu}$ is
parameterization independent, so that the product
\begin{equation}
\scrD f^1 \scrD f^2
\Det\left[ \tilde t^{\mu\nu} \frac{\partial x_p^\mu}
{\partial f^1} \frac{\partial x_p^\nu} {\partial f^2} \right]
\end{equation}
is independent of $\tilde\xi(\xi)$.
Therefore, this product, along with $J_\parallel$ and $e^{iS_{{\rm eff}}}$,
can be brought outside the $\tilde\xi$ integral in
\eqnlessref{W subbed}. The path integral then takes the form
\begin{equation}
W[\Gamma] = \int \scrD f^1 \scrD f^2
\Det\left[ \tilde t^{\mu\nu} \frac{\partial x_p^\mu}
{\partial f^1} \frac{\partial x_p^\nu} {\partial f^2} \right]
J_\parallel e^{iS_{{\rm eff}}} \int \scrD \tilde\xi
\Det[\sqrt{-g^p}] \,.
\label{before cancel J parallel}
\end{equation}

The remaining integral over reparameterizations $\tilde\xi$
is equal to $J_\parallel^{-1}$, defined by \eqnlessref{Jparallel},
and is canceled by the explicit factor of $J_\parallel$ appearing
in \eqnlessref{before cancel J parallel}. This means we
do not need to evaluate $J_\parallel$, and can
avoid the complications inherent in evaluating the
integral over reparameterizations of the string
coordinates. The anomalies produced in string theory
by evaluating this integral are not present, so we do
not have a Polchinski--Strominger term in the theory.
\eqnref{before cancel J parallel} gives the final result for the
Wilson loop
\begin{equation}
W[\Gamma] = \int \scrD f^1 \scrD f^2 \Det\left[ \tilde t_{\mu\nu}
\frac{\partial x_p^\mu}{\partial f^1} \frac{\partial x_p^\nu}
{\partial f^2} \right] e^{iS_{{\rm eff}}} \,,
\label{param measure}
\end{equation}
as an integration over two function $f^1(\xi)$ and $f^2(\xi)$,
the physical degrees of freedom of the string.

The path integral \eqnlessref{param measure} is invariant
under reparameterizations
of the string, and describes a two dimensional field
theory with two degrees of freedom, the two transverse
oscillations of a two dimensional sheet.
The integration \eqnlessref{param measure} goes over
the normal fluctuations of the string worldsheet.
The components of $f^1$ and $f^2$ along the sheet
are nonphysical. The determinant
in \eqnlessref{param measure} is a normalization factor for
$f^1$ and $f^2$. This can be seen by applying
the identity $\tilde t_{\mu\nu} =
\epsilon^{AB} n_{\mu A} n_{\nu B}$ to the determinant,
\begin{equation}
\Det\left[ \tilde t_{\mu\nu}
\frac{\partial x_p^\mu}{\partial f^1} \frac{\partial x_p^\nu}
{\partial f^2} \right]
= \Det\left[ \epsilon^{AB} \left( n_{\mu A} \frac{\partial x_p^\mu}
{\partial f^1} \right) \left( n_{\nu B} \frac{\partial x_p^\nu}
{\partial f^2} \right) \right] \,.
\end{equation}
The factors of $n_{\mu A} \frac{\partial x_p^\mu}
{\partial f^i}$ determine the amount of the fluctuation
$f^i$ which is in a direction normal to the sheet.

\eqnref{param measure} is the string representation of
any field theory containing classical vortex solutions.
The expression \eqnlessref{Seff} for
$S_{{\rm eff}}$ had been obtained previously by
Gervais and Sakita~\cite{Gervais+Sakita}; we are unaware of
any previous derivation of the string
representation \eqnlessref{param measure}
for the path integral. We now show how it provides a method for
explicit calculations.

\section{The Semiclassical Approximation}

\label{small f section}

In this section we carry out the semiclassical expansion
of $W[\Gamma]$ about a classical solution of the
effective string theory, and find the leading contribution
of string fluctuations to the effective action $-i\log W[\Gamma]$.
The ends of the string follow the path $\Gamma$ fixed by the
prescribed trajectory of the quarks, and the fluctuations of
the string are cutoff at the momentum scale $M$ of the inverse
string radius. As explained in section~\ref{effective action section},
we take the action $S_{{\rm eff}}$ of the effective
string theory to be the Nambu--Goto action
\begin{equation}
S_{{\rm eff}} = -\sigma \int d^2\xi \sqrt{-g} \,,
\label{NG action again}
\end{equation}
and \eqnlessref{param measure} becomes
\begin{equation}
W[\Gamma] = \int \scrD f^1(\xi) \scrD f^2(\xi)
\Det\left[ \tilde t^{\mu\nu} \frac{\partial x_p^\mu}{\partial f^1}
\frac{\partial x_p^\nu}{\partial f^2} \right]
e^{-i\sigma \int d^2\xi \sqrt{-g}} \,.
\label{W with action}
\end{equation}

We expand \eqnlessref{W with action} in small
fluctuations $f^i$ of $x_p^\mu[f^i,\xi]$ around a
fixed sheet $\bar x_p^\mu(\xi)$, subject to the condition that
the boundary of $\bar x_p^\mu$ lies on the curve $\Gamma$,
\begin{equation}
x_p^\mu(f^i,\xi) = \bar x_p^\mu(\xi) + f^i \frac{\partial x_p^\mu}
{\partial f^i} \Bigg|_{f^i=0}
+ \frac{1}{2} f^i f^j \frac{\partial^2 x_p^\mu}
{\partial f^i \partial f^j} \Bigg|_{f^i=0} + O(f^3) \,,
\end{equation}
where $\bar x_p^\mu(\xi) \equiv x_p^\mu(f^i=0,\xi)$ is the
position of the string worldsheet when $f^1=f^2=0$. Expanding $\sqrt{-g}$
to quadratic order in small $f^1$ and $f^2$, we obtain
\begin{eqnarray}
W[\Gamma] &=& \int \scrD f^i(\xi) \Det\left[ \frac{1}{2}
\tilde t^{\mu\nu} \epsilon^{ij} \frac{\partial x_p^\mu}{\partial f^i}
\frac{\partial x_p^\nu}{\partial f^j} \Bigg|_{f^i=0} \right]
\nonumber \\
& & \times \exp\left\{ -i\sigma \int d^2\xi \sqrt{-\bar g} \left[ 1
+ \bar g^{ab} \frac{\partial x_p^\mu}{\partial\xi^a} \Bigg|_{f^i=0}
\frac{\partial}{\partial\xi^b} \left( \frac{\partial x_p^\mu}
{\partial f^i} \Bigg|_{f^i=0} f^i \right)
+ \frac{1}{2} f^i G^{-1}_{ij} f^j \right] \right\} \,,
\label{semiclassical string partition}
\end{eqnarray}
where $\bar g_{ab}$ is
\begin{equation}
\bar g_{ab} = \frac{\partial \bar x_p^\mu}{\partial\xi^a}
\frac{\partial \bar x_p^\mu}{\partial\xi^b} \,,
\end{equation}
the metric of the fixed worldsheet $\bar x_p^\mu$, and
\begin{equation}
G^{-1}_{ij} = \frac{1}{\sqrt{-g}}
\frac{\partial^2 \sqrt{-g}}{\partial f^i \partial f^j} \Bigg|_{f^i=0} \,.
\label{G^-1_ij def}
\end{equation}

We choose $\bar x_p^\mu$ to be the surface which minimizes
the action. Then $\bar x_p^\mu$ satisfies the
``classical equation of motion''
\begin{equation}
\frac{\partial x_p^\mu}{\partial f^i} \Bigg|_{f^i=0} (-\nabla^2)
\bar x_p^\mu = 0 \,,
\label{classical string equation}
\end{equation}
where the covariant Laplacian is
\begin{equation}
-\nabla^2 = \frac{1}{\sqrt{-\bar g}} \frac{\partial}{\partial\xi^a}
\bar g^{ab} \sqrt{-\bar g} \frac{\partial}{\partial\xi^b} \,.
\end{equation}
Using the fact that the covariant derivative of the
metric is zero, we show in Appendix~\ref{notation appendix}
that
\begin{equation}
(\partial_a \bar x_p^\mu)(-\nabla^2 \bar x_p^\mu) = 0 \,.
\label{covariant identity}
\end{equation}
The vectors $\frac{\partial x_p^\mu}{\partial f^i} \big|_{f^i=0}$
and $\partial_a \bar x_p^\mu$ form a complete basis,
so \eqnlessref{classical string equation}
and \eqnlessref{covariant identity} imply
\begin{equation}
-\nabla^2 \bar x_p^\mu = 0 \,.
\label{class string simple}
\end{equation}

Evaluating the $f^i$ integral in \eqnref{semiclassical string partition}
gives
\begin{equation}
W[\Gamma] = e^{-i\sigma \int d^2\xi \sqrt{-\bar g}}
\Det\left[ \frac{1}{2} \tilde t_{\mu\nu}
\epsilon^{ij} \frac{\partial x_p^\mu}{\partial f^i}
\frac{\partial x_p^\nu}{\partial f^j} \Bigg|_{f^i=0} \right]
\Det^{-1/2}\left[G^{-1}_{ij}\right] \,.
\label{Z of G}
\end{equation}
The inverse propagator $G^{-1}_{ij}$ \eqnlessref{G^-1_ij def}
can be shown to be
\begin{equation}
G^{-1}_{ij} = - \frac{\partial x_p^\mu}{\partial f^i} \Bigg|_{f^i=0}
\bar n_{\mu A} \left[ -\nabla^2 \delta_{AB}
- \bar\scrK^A_{ab} \bar\scrK^{Bab} \right]
\bar n_{\nu B} \frac{\partial x_p^\nu}{\partial f^j} \Bigg|_{f^i=0} \,,
\label{G inverse def}
\end{equation}
where $\bar\scrK^A_{ab}$ is the extrinsic curvature tensor
of the sheet $\bar x_p^\mu$. A derivation of \eqnlessref{G inverse def}
is given in appendix A of \cite{Luscher1}. The $\bar n_{\mu A}$ are vectors
normal to the worldsheet $\bar x_p^\mu$.
\eqnref{G inverse def} gives
\begin{equation}
\Det^{-1/2}\left[G^{-1}_{ij}\right] = \Det^{-1/2}\left[ -\nabla^2 \delta_{AB}
- \bar\scrK^A_{ab} \bar\scrK^{Bab} \right]
\Det^{-1}\left[\frac{1}{2} \epsilon^{AB} \bar n_{\mu A} \bar n_{\nu B}
\epsilon^{ij} \frac{\partial x_p^\mu}{\partial f^i}
\frac{\partial x_p^\nu}{\partial f^j} \Bigg|_{f^i=0} \right] \,.
\label{G det}
\end{equation}
From the identity \eqnlessref{dual normal},
\begin{equation}
\tilde t_{\mu\nu} = \epsilon^{AB} \bar n_{\mu A} \bar n_{\nu B} \,,
\end{equation}
we see that the first determinant in \eqnlessref{Z of G} and the second
determinant in \eqnlessref{G det} cancel. The determinant
appearing in \eqnlessref{param measure}
produces exactly the correct normalization
for the Green's function. The functional integral \eqnlessref{Z of G}
becomes
\begin{equation}
W[\Gamma] = e^{-i\sigma \int d^2 \xi \sqrt{-\bar g}}
\Det^{-1/2}\left[ -\nabla^2 \delta_{AB}
- \bar\scrK^A_{ab} \bar\scrK^{Bab} \right] \,.
\label{last Z}
\end{equation}

We note that \eqnlessref{last Z} is independent of the factors of $n_\mu^A
\frac{\partial x_p^\mu}{\partial f^i} \big|_{f^i=0}$ which appeared
in the inverse propagator \eqnlessref{G inverse def}.
These factors are the projections of the fluctuations $f^i$
normal to the string worldsheet. For small
$f^i$, the worldsheet $x_p^\mu$ is
\begin{equation}
x_p^\mu = \bar x_p^\mu + f_i \frac{\partial x_p^\mu}{\partial f^i}
\Bigg|_{f^i=0} + O({f^i}^2) \,.
\end{equation}
The perturbation of the worldsheet in the direction $\bar n^{\mu A}$ is
\begin{equation}
n_\mu^A \left( x_p^\mu - \bar x_p^\mu \right)
= n_\mu^A \frac{\partial\bar x_p^\mu}{\partial f^i} \Bigg|_{f^i=0} f_i
+ O({f^i}^2) \,.
\end{equation}
The factors of $n_\mu^A \frac{\partial x_p^\mu}{\partial f^i}
\big|_{f^i=0}$ in \eqnlessref{G inverse def}
project out the part of the fluctuations $f^i$
perpendicular to the worldsheet $\bar x_p^\mu$.
Only normal fluctuations contribute to $W[\Gamma]$,
since fluctuations along the worldsheet are equivalent
to a reparameterization of the sheet coordinates.

The effective action obtained from \eqnlessref{last Z} is
\begin{equation}
-i\ln W[\Gamma] = S_{cl} + S_{{\rm fluc}} \,.
\label{effective action def}
\end{equation}
The first term in \eqnlessref{effective action def}
is the Nambu--Goto action
evaluated at the ``classical'' worldsheet $\bar x_p^\mu(\xi)$,
\begin{equation}
S_{cl} = -\sigma \int d^2\xi \sqrt{-\bar g} \,.
\label{S_NG def}
\end{equation}
The semiclassical correction $S_{{\rm fluc}}$ due to the transverse
string fluctuations is
\begin{equation}
S_{{\rm fluc}} = \frac{i}{2} \Tr\ln\left[ -\nabla^2 \delta_{AB}
- \bar \scrK^A_{ab} \bar \scrK^{Bab} \right] \,.
\label{W unreg}
\end{equation}
To summarize, we have integrated out the string fluctuations,
and reduced the problem to the evaluation of the determinant in
\eqnlessref{last Z}. This is a quantum mechanical scattering problem
in the background of the solution of the classical equation
\eqnlessref{class string simple}, with appropriate
boundary conditions. In the next section, we describe how
to evaluate this determinant.

\section{Regularization of String Integrals
and the Role of the L\"uscher Term}

\label{regularization section}

The argument of the logarithm in \eqnlessref{W unreg}
is the inverse propagator for fluctuations
on the string. This inverse propagator can also
be obtained by direct
variation of the Nambu--Goto action with respect
to any transverse coordinates $x_\perp^A$,
\begin{equation}
\frac{\partial^2 S}{\partial x_\perp^A \partial x_\perp^B}
= -\nabla^2 \delta_{AB} - \bar \scrK^A_{ab} \bar \scrK^{Bab} \,,
\end{equation}
up to an overall normalization factor. In fact, the correction
\eqnlessref{W unreg} to the effective action has already been studied by
L\"uscher, Symanzik, and Weisz (LSW)~\cite{Luscher1} in the case of
a straight string with fixed ends. We describe their results, which we
will use in evaluating \eqnlessref{W unreg}.
LSW used Pauli--Villars regularization to obtain a
regulated form $S_{{\rm reg}}$ of the trace in \eqnlessref{W unreg},
\begin{eqnarray}
S_{{\rm reg}} = - \int_0^\infty \frac{dt}{t}
\left(1 + \sum_j \epsilon_j e^{-t{\cal M}_j^2}\right) \Tr \,
e^{-t \left(-\nabla^2 \delta_{AB}
- \bar \scrK^A_{ab} \bar \scrK^{Bab}\right)} \,.
\label{regulated action}
\end{eqnarray}
The ${\cal M}_j$ are the masses of the regulators, and the $\epsilon_j$ are
suitably chosen coefficients. The Laplacian in
\eqnlessref{regulated action} has been
Wick rotated from Minkowski to Euclidean space.

The regulated quantity $S_{{\rm reg}}$ is separated into
a divergent part $S_{{\rm div}}$ and a finite part $S_{{\rm PV}}$,
\begin{equation}
S_{{\rm reg}} = S_{{\rm div}}({\cal M}_j,\epsilon_j)
+ S_{{\rm PV}} \,.
\end{equation}
LSW evaluated the divergent part
$S_{{\rm div}}({\cal M}_j,\epsilon_j)$, and obtained
terms which are quadratically, linearly, and logarithmically
divergent in the cutoffs ${\cal M}_j$. The quadratic term is a renormalization
of the string tension, the linear term is a renormalization
of the quark masses, and the logarithmically divergent term
is proportional to the integral over all space of the
scalar curvature ${\cal R}$ of the string worldsheet.

LSW also obtained a formal expression for the finite part
$S_{{\rm PV}}$. They evaluated this expression only
for the case of a straight string of length $R$ with fixed ends,
and calculated a correction $V_{\hbox{\scriptsize L\"uscher}}$
to the static potential:
\begin{equation}
V_{\hbox{\scriptsize L\"uscher}} = - \lim_{T\to\infty} \frac{1}{T}
S_{{\rm PV}} = -\frac{\pi}{12 R} \,.
\label{luscher result}
\end{equation}

We are interested in calculating $S_{{\rm fluc}}$ for rotating quarks,
so we must evaluate $S_{{\rm reg}}$ for a more general
surface. We break \eqnlessref{W unreg}
into two parts:
\begin{equation}
S_{{\rm reg}} = i\Tr\ln[-\nabla^2]
+ \frac{i}{2} \Tr\ln\left[ \frac{-\nabla^2 \delta_{AB}
- \bar \scrK^A_{ab} \bar \scrK^{Bab}}{-\nabla^2} \right] \,.
\label{separated effective action}
\end{equation}
We will evaluate the first term in \eqnlessref{separated effective action}
by generalizing the calculation of LSW.
We will calculate the second term directly.

The first term in \eqnlessref{separated effective action},
\begin{equation}
S_1 = i\Tr\ln[-\nabla^2] \,,
\label{W_1}
\end{equation}
involves the Laplacian in the curved background of the
classical solution $\bar x_p^\mu$. In the
flat case studied by LSW, the Laplacian is equal to $-\partial^2$,
\begin{equation}
-\partial^2 = \frac{\partial^2}{\partial t^2}
- \frac{\partial^2}{\partial r^2} \,.
\label{flat sheet Laplacian}
\end{equation}
The coordinate $t$ is the time in the lab frame, and $r$ is a radial
coordinate which takes the values $-R_1$ and $R_2$ at the
two ends of the string. The length of the string is
$R = R_1 + R_2$.

To calculate $S_1$ we extend the calculation of LSW to
more general coordinate systems. We make a coordinate transformation
$\xi \to \xi'$ to conformal coordinates, where the
transformed metric $g_{ab}' = \eta_{ab} e^\varphi$,
$\eta_{ab} = {\rm diag}(-1,1)$.
This transformation puts the Laplacian in
a form similar to the flat sheet Laplacian
\eqnlessref{flat sheet Laplacian}, and allows us
to evaluate \eqnlessref{W_1} by extending the calculation
of LSW.

To see how this works, we express $e^{iS_1}$ as a functional integral,
\begin{equation}
e^{iS_1} = \int \scrD f_1 \scrD f_2 \exp\left\{-i\int d^2\xi \sqrt{-g} g^{ab}
\frac{\partial f^i}{\partial\xi^a} \frac{\partial f^i}{\partial\xi^b}
\right\} \,.
\label{W_1 int}
\end{equation}
The transformation to conformal coordinates $\xi'$ gives
\begin{equation}
e^{iS_1} = \int \scrD f_1 \scrD f_2 \exp\left\{-i\int d^2\xi' \eta^{ab}
\frac{\partial f^i}{\partial{\xi'}^a} \frac{\partial f^i}{\partial{\xi'}^b}
\right\} \,,
\label{W_1 conformal}
\end{equation}
which is of the form of $S_{{\rm reg}}$ treated by LSW.

We will need $S_1$ in the limit of large $T$, and hence
are only interested in strings whose metric is time independent.
To determine which metrics are time independent, we must choose
a coordinate system. We choose coordinates
$r$ and $t$, where $t$ is the time in the lab frame,
and $r$ is orthogonal to $t$ ($g_{rt} = 0$).
This guarantees that $t$ is the physical time. From
now on we consider only metrics
$g_{ab}$ which are independent of $t$.

In Appendix \ref{luscher appendix}, we show that
\begin{equation}
S_1 = S_{1,{\rm div}} + S_{1,{\rm finite}} \,,
\end{equation}
where $S_{1,{\rm div}}$ contains quark mass and string tension
renormalizations. The finite part of $S_1$ is
\begin{equation}
S_{1,{\rm finite}} = T \frac{\pi}{12 R_p} \,,
\label{effective R proper}
\end{equation}
where
\begin{equation}
R_p = \int_{-R_1}^{R_2} dr \sqrt{\frac{\bar g_{rr}(r)}{-\bar g_{tt}(r)}} \,.
\label{R_p def text}
\end{equation}
The results \eqnlessref{effective R proper} and
\eqnlessref{R_p def text} are valid for any
orthogonal coordinate system with a time independent metric.

We show in Appendix~\ref{luscher appendix} that
$R_p$ is equal to the classical energy of the
string divided by the string tension $\sigma$.
We call $R_p$ the ``proper length'' of the string.
For a flat metric, where $\bar g_{rr} = - \bar g_{tt} = 1$, the proper
length $R_p$ of the string reduces to the distance $R$
between its endpoints.

\section{Correction to the Effective Action for a Curved Sheet}

\label{W_2 section}

In the previous section, we evaluated the finite part of $S_1$ for
a sheet with a time independent metric using the results of LSW.
In this section
we evaluate the second term in \eqnlessref{separated effective action},
\begin{equation}
S_2 \equiv i\frac{T}{2} \Tr\ln\left[\frac{-\nabla^2 \delta_{AB}
- \bar\scrK^A_{ab} \bar\scrK^{Bab}}{-\nabla^2}\right] \,.
\label{W_2 def}
\end{equation}
The trace in \eqnlessref{W_2 def} is over functions of
$r$ and $t$.

We first make the coordinate transformation $r \to x$,
\begin{equation}
\frac{dx}{dr} = \sqrt{\frac{\bar g_{rr}(r)}{-\bar g_{tt}(r)}}
\,, \kern 1 in x\big|_{r=0} = 0 \,.
\label{coord x def}
\end{equation}
The coordinate $x$ runs from $-X_1$ to $X_2$,
\begin{eqnarray}
X_1 &=& \int_{-R_1}^0 dr \sqrt{\frac{\bar g_{rr}(r)}{-\bar g_{tt}(r)}}
\nonumber \\
X_2 &=& \int_0^{R_2} dr \sqrt{\frac{\bar g_{rr}(r)}{-\bar g_{tt}(r)}} \,,
\end{eqnarray}
and $X_1 + X_2 = R_p$.
In Appendix~\ref{luscher appendix}, we show that the metric in the
system $(x,t)$ is conformal ($\bar g_{xx} = - \bar g_{tt}$,
$\bar g_{xt} = 0$). In this coordinate system, the inverse
propagator for string fluctuations is
\begin{equation}
-\nabla^2 \delta_{AB} - \bar\scrK^A_{ab} \bar\scrK^{Bab}
= \frac{1}{\sqrt{-\bar g}} \left(\frac{\partial^2}{\partial t^2}
- \frac{\partial^2}{\partial x^2}\right) \delta_{AB}
- \bar\scrK^A_{ab} \bar\scrK^{Bab} \,.
\end{equation}
The string has infinite extent in time, and the curvature
$\bar\scrK^A_{ab} \bar\scrK^{Bab}$ is independent of $t$,
so we can take the Fourier transform with respect
to the time coordinate. We
express the trace in \eqnlessref{W_2 def} over functions of $t$ and $r$
as an integral over a frequency $\nu$ and a trace over functions of
a single variable $x$,
\begin{equation}
S_2 = \frac{T}{2} \int_{-\infty}^\infty \frac{d\nu}{2\pi} \Tr_x
\ln\left[\frac{\left(\nu^2 - \frac{\partial^2}{\partial x^2}\right)
\delta_{AB} - \sqrt{-\bar g} \bar\scrK^A_{ab} \bar\scrK^{Bab}}
{\nu^2 - \frac{\partial^2}{\partial x^2}}\right] \,.
\label{W_2 of nu}
\end{equation}
In going from \eqnlessref{W_2 def} to \eqnlessref{W_2 of nu}, we
have also carried out the Wick rotation $\nu \to -i\nu$. The
integration over $\nu$ gives
\begin{equation}
S_2 = \frac{T}{2} \left[
\Tr_x \sqrt{-\frac{\partial^2}{\partial x^2}\delta_{AB}
- \sqrt{-\bar g} \bar\scrK^A_{ab} \bar\scrK^{Bab}}
- \Tr\sqrt{-\frac{\partial^2}{\partial x^2}\delta_{AB}} \right] \,.
\label{W_2 given}
\end{equation}
\eqnref{W_2 given} expresses $S_2$ as the trace of the difference of
two operators. The first has the form of a Hamiltonian for a
relativistic particle in the local potential
$\sqrt{-\bar g} \bar\scrK^A_{ab} \bar\scrK^{Bab}$.
The second operator has the form of a
free Hamiltonian. The square roots enter because we are
working with relativistic degrees of freedom.

The terms $S_1$ and $S_2$ are proportional to the
time $T$, but are otherwise time independent.
We define the ``effective Lagrangian'' of the
string to be the effective action divided by the time $T$.
The sum of \eqnlessref{effective R proper}
and \eqnlessref{W_2 given} gives the
effective Lagrangian $L_{{\rm fluc}}$ determining the
contribution of the string fluctuations to $W[\Gamma]$,
\begin{eqnarray}
L_{{\rm fluc}} &\equiv& \lim_{T\to\infty} \frac{1}{T}
\left( S_{1,{\rm finite}} + S_2 \right)
\nonumber \\
&=& \frac{\pi}{12 R_p}
+ \frac{1}{2} \left( \Tr\sqrt{-\frac{\partial^2}{\partial x^2}
\delta_{AB} - \sqrt{-\bar g} \bar\scrK^A_{ab} \bar\scrK^{Bab}}
- \Tr\sqrt{-\frac{\partial^2}{\partial x^2} \delta_{AB}} \right) \,.
\label{rotating potential}
\end{eqnarray}
In the next section, we will evaluate $L_{{\rm fluc}}$ for a
string of length $R$ rotating with angular velocity $\omega$.

In Appendix~\ref{log cutoff appendix}, we show that,
for a general sheet, $S_2$ is logarithmically divergent.
We show that its divergent part is given by
\begin{equation}
S_{2,{\rm div}} = \frac{T}{4} \sum_{n=1}^{M R_p/\pi} \frac{1}{\pi n}
\int_{-X_1}^{X_2} dx \sqrt{-\bar g} {\cal R} \,,
\end{equation}
where ${\cal R}$ is the scalar curvature,
\begin{equation}
{\cal R} = \left( \scrK^{Aa}_a \right)^2
- \left( \scrK^A_{ab} \right)^2 \,.
\end{equation}
This result agrees, in the large time limit, with
the logarithmically divergent term in the
cutoff dependent part of the effective
string action \eqnlessref{Luscher reg}
found by LSW.

\section{Effective Lagrangian for Rotating String}

\label{regge effective potential section}

We now evaluate the effective Lagrangian
of a string with boundary $\Gamma$
generated by a quark--antiquark pair
separated at fixed distances $R_1$ and $R_2$
from the origin, and rotating with angular
velocity $\omega$. This Lagrangian has two parts, the
classical string Lagrangian
and the contribution \eqnlessref{rotating potential}
of string fluctuations.

We evaluate the classical string Lagrangian first.
The solution to the classical equations of
motion \eqnlessref{class string simple} yields the classical,
straight rotating string,
\begin{equation}
\bar x^\mu(r,t) = t \ehat_0^\mu + r \cos(\omega t) \ehat_1^\mu
+ r \sin(\omega t) \ehat_2^\mu \,.
\label{x bar def}
\end{equation}
The coordinate $r$ is chosen so that the velocity of
the string is zero when $r = 0$.
The coordinate $r$ runs from $-R_1$ to $R_2$, and $t$ runs
from $-\infty$ to $\infty$. The vectors $\ehat_1^\mu$ and $\ehat_2^\mu$
are two orthogonal unit vectors in the plane of rotation,
and $\ehat_0^\mu$ is a unit vector in the time
direction. The classical Lagrangian $L_{cl}^{{\rm string}}$ obtained from
\eqnlessref{S_NG def} is
\begin{eqnarray}
L_{cl}^{{\rm string}} &\equiv& -\lim_{T\to\infty} \frac{1}{T}
\sigma \int d^2\xi \sqrt{-\bar g}
\nonumber \\
&=& -\sigma \int_{-R_1}^{R_2} dr \sqrt{1-r^2\omega^2}
\nonumber \\
&=& -\sigma \sum_{i=1}^2 \frac{R_i}{2} \left( \frac{\arcsin (\omega R_i)}
{\omega R_i} + \sqrt{1 - R_i^2 \omega^2} \right) \,.
\label{L classical string}
\end{eqnarray}

Next, we calculate the contribution $L_{{\rm fluc}}$
\eqnlessref{rotating potential} due to string fluctuations.
The metric of the sheet \eqnlessref{x bar def} is
\begin{equation}
\bar g_{tt} = -1 + r^2\bar\omega^2 \,,\,\,\,\,\,
\bar g_{rr} = 1 \,,\,\,\,\,\,
\bar g_{rt} = 0 \,.
\label{r,t metric def}
\end{equation}
This metric is independent of $t$, and $\bar g_{rt} = 0$.
We make the transformation \eqnlessref{coord x def}
from coordinates $r$ and $t$
to coordinates $x$ and $t$, and find
\begin{equation}
x = \frac{1}{\omega} \arcsin\omega r \,.
\end{equation}
The coordinate $x$ runs from $-X_1$ to $X_2$, where
\begin{equation}
X_i = \frac{\arcsin \left(\omega R_i\right)}{\omega} \,,
\end{equation}
and the proper length $R_p$ of the string is
\begin{equation}
R_p = X_1 + X_2 = \sum_i \frac{\arcsin\left(\omega R_i\right)}{\omega} \,.
\end{equation}

Using this coordinate system , we evaluate $L_{{\rm fluc}}$ in
Appendix~\ref{eigenfunctions and sum} for the case
of equal quark masses, $R_1 = R_2 = R/2$:
\begin{equation}
L_{{\rm fluc}} = \frac{v}{\arcsin v} \frac{\pi}{12R}
- \frac{2 v}{\pi R} \left[ v \gamma
\ln\left(\frac{M R}{2(\gamma^2 - 1)}\right)
+ v \gamma - \frac{\pi}{2} \right]
- \frac{v^2 \gamma}{\pi R} f\left(v\right) \,,
\label{V string}
\end{equation}
where
\begin{equation}
v \equiv \frac{R}{2} \omega \,, \kern 1 in
\gamma = \frac{1}{\sqrt{1-v^2}} \,,
\label{v and gamma def}
\end{equation}
and the function $f(v)$ is
\begin{equation}
f(v) = \int_0^\infty ds \ln\left[\frac{s^2 + 2s \coth\left(
2s v \gamma \arcsin v \right) + 1}{(s+1)^2}\right] \,.
\end{equation}
\eqnref{V string} becomes
the L\"uscher term in the zero velocity limit.

We are interested in the large $R$ limit, where
the quark velocity is close to the speed of light.
For $v$ close to one, \eqnref{V string} becomes
\begin{equation}
L_{{\rm fluc}} = - \frac{2}{\pi R} \gamma \left[
\ln\left( \frac{M R}{2\gamma^2} \right) + 1\right]
+ \frac{7}{6R} + O\left(\frac{\ln\gamma}{\gamma R}\right) \,.
\label{L fluc calculated}
\end{equation}

Furthermore, for the semiclassical expansion to be valid, the
theory must be weakly coupled. That is, $L_{{\rm fluc}}$
must be less than $L_{cl}^{{\rm string}}$
\eqnlessref{L classical string}. For large $R$,
\begin{equation}
L_{cl}^{{\rm string}} = - \frac{\pi}{4} \sigma R  -  \frac{\pi}{8} \sigma R
\gamma^{-2} +  \frac{1}{6} \sigma R \gamma^{-3}
+ O(\gamma^{-4} \sigma R) \,.
\end{equation}
The semiclassical expansion is valid, since,
as we will see, $R$ grows like $\gamma^2$ in
the $v \to 1$ limit. In this case, the long distance limit
where the effective theory is applicable is automatically the
region of weak coupling.

\section{Regge Trajectories}

\label{regge plot section}

We calculate classical Regge trajectories for equal
mass quarks by adding a quark mass
term to the string Lagrangian $L_{cl}^{{\rm string}}$,
\begin{equation}
L_{cl} = L_{cl}^{{\rm string}} - 2m\sqrt{1-v^2} = -\sigma \frac{R}{2}
\left(\frac{\arcsin v}{v} + \gamma^{-1}\right)
- 2m\gamma^{-1} \,.
\label{L_cl def}
\end{equation}
We have used \eqnref{L classical string} with $R_1 = R_2 = R/2$.
The quark velocity is $v = \omega R/2$, and $\gamma = 1/\sqrt{1-v^2}$
is the quark boost factor. The Lagrangian \eqnlessref{L_cl def}
is a function of $R$ and $\omega$,
\begin{equation}
L_{cl} = L_{cl}(R,\omega) \,.
\end{equation}
The angular momentum of the meson is obtained by varying the Lagrangian
with respect to the angular velocity,
\begin{equation}
J = \frac{\partial L_{cl}}{\partial\omega}
= \sigma \frac{R^2}{4v} \left( \frac{\arcsin v}{v}
- \gamma^{-1} \right) + m R v \gamma^{-1} \,.
\label{J equation}
\end{equation}
The meson energy is given by the Hamiltonian,
\begin{equation}
E = \omega \frac{\partial L_{cl}}{\partial\omega} - L_{cl}
= \sigma R \frac{\arcsin v}{v} + 2m\gamma \,.
\label{E equation}
\end{equation}

The classical equation of motion
\begin{equation}
\frac{\partial L}{\partial R} = 0
\end{equation}
for the quarks determines $R$ as a function of $\omega$,
\begin{equation}
\sigma \frac{R}{2} = m (\gamma^2 - 1) \,.
\label{boundary condition}
\end{equation}
\eqnref{boundary condition} shows that $R$ is proportional
to $\gamma^2$ for large $\gamma$. Expanding \eqnlessref{J equation}
and \eqnlessref{E equation} in the large $R$ limit, where
the quark velocity $v$ goes to one, yields the result:
\begin{equation}
\frac{J}{E_{cl}^2} = \frac{1}{2\pi\sigma} \left( 1 - \frac{8}{3\pi}
\gamma^{-3} + O(\gamma^{-5}) \right) \,.
\label{classical Regge eqn}
\end{equation}
The first term in \eqnlessref{classical Regge eqn} is
the classical formula for the slope of a Regge trajectory. The second
term is the leading correction for nonzero classical quark mass, where
$\gamma^{-1} = \sqrt{1-v^2} \ne 0$.

We now calculate the correction to the energy obtained by
considering $L_{{\rm fluc}}$ a small perturbation to
the classical Lagrangian $L_{cl}$. The Lagrangian,
\begin{equation}
L(\omega) = L_{cl}(\omega) + L_{{\rm fluc}}(\omega) \,,
\end{equation}
depends on only one degree of freedom, the rotation
angle $\theta$, through its time derivative $\omega = \dot\theta$,
To first order in $L_{{\rm fluc}}$, the correction to the energy is minus
the correction to the Lagrangian~\cite{Landau+Lifshitz:Mechanics},
\begin{equation}
E(J) = \left[E_{cl}(\omega) - L_{{\rm fluc}}(\omega)
\right] \Bigg|_{\omega = \omega(J)} \,,
\label{E of J}
\end{equation}
where $\omega$ is given as a function of $J$ through the
classical relation \eqnlessref{J equation}.

The correction \eqnlessref{E of J} to the energy of the meson
gives a correction to the slope \eqnlessref{classical Regge eqn}
of a Regge trajectory,
\begin{equation}
\frac{J}{E^2} = \frac{J}{E_{cl}^2} \frac{E_{cl}^2}{E^2}
\simeq \frac{J}{E_{cl}^2} \left( 1 + 2 \frac{L_{{\rm fluc}}}{E} \right) \,.
\end{equation}
Using \eqnlessref{classical Regge eqn} for $J/E_{cl}^2$
and \eqnlessref{L fluc calculated} for $L_{{\rm fluc}}$,
we obtain
\begin{equation}
\frac{J}{E^2} = \frac{1}{2\pi\sigma}
- \frac{2}{\pi^2 \sigma R E} \gamma \left[
\ln\left( \frac{M R} {2\gamma^2} \right) + 1\right]
- \frac{4}{3\pi^2\sigma} \gamma^{-3} + \frac{7}{6 \pi \sigma R E}
+ O\left(\gamma^{-5}, \frac{1}{RE\gamma}\right) \,.
\label{Regge corrected}
\end{equation}
We write $R$ and $\gamma$ as functions of $E$ using
the definition \eqnlessref{E equation} of $E$ and the
classical equation of motion \eqnlessref{boundary condition}.
Because $R$ and $\gamma$ only appear in the small correction
terms in the result \eqnlessref{Regge corrected}, we only need
their leading order dependence on $E$,
\begin{equation}
R \simeq \frac{2E}{\pi\sigma} \,,
\kern 1 in
\gamma \simeq \sqrt{\frac{E}{\pi m}} \,.
\label{R,gamma of E}
\end{equation}
Substituting \eqnlessref{R,gamma of E} in \eqnlessref{Regge corrected}
gives
\begin{equation}
J = \frac{E^2}{2\pi\sigma} - \sqrt{\frac{E}{\pi^3 m}} \left[
\ln\left(\frac{Mm}{\sigma} \right) + 1 \right]
- \frac{4}{3\sigma} \sqrt{\frac{m^3 E}{\pi}} + \frac{7}{12}
+ O\left( E^{-1/2} \right) \,.
\label{J result}
\end{equation}
The leading term is the classical Regge formula.
The next term is the leading correction due to string fluctuations.
The third term is a nonzero quark mass correction. The fourth term
is another correction due to string fluctuations.

\eqnref{J result} gives a meson Regge trajectory $J(E^2)$.
\begin {figure}[ht]
    \begin{center}
	\begin{tabular}{rc}
	    \vbox{\hbox{$J$ \hskip 0.15in \null} \vskip 1.25in} &
	    \epsfbox{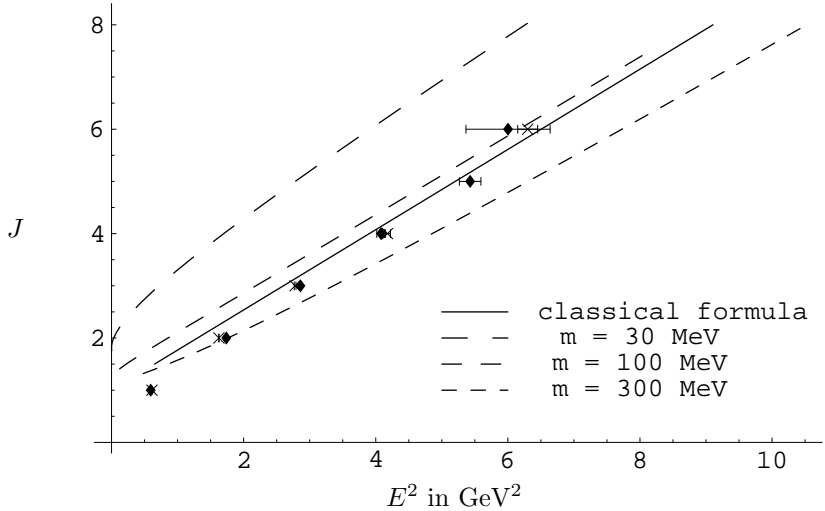} \\
	    &
	    \hbox{$E^2$ in GeV$^2$} \\
	\end{tabular}
    \end{center}
    \medskip
    \caption{$J$ versus $E^2$}
\label{regge figure}
\end {figure}
We used values $\frac{4}{3} \alpha_s = 0.25$
and $\sigma = (455 \hbox{ MeV})^2$ obtained from the
Cornell fits of heavy quark potentials~\cite{Eichten+Gottfried}.
This gives $M = g\phi_0 = \sqrt{\frac{3\sigma}{4\alpha_s}}
= 910 \hbox{ MeV}$. The only other parameter is the quark mass $m$.
In Figure~\ref{regge figure}, we plot $J$
versus the square of the energy \eqnlessref{E of J}
for quark masses of 30 MeV, 100 MeV, and 300 MeV. For comparison,
we also plot the classical formula $J = E^2/2\pi\sigma$.
The points~\cite{data_book}
plotted on the graph are the $\rho^1$, $A^2$, $\rho^3$, $A^4$, $\rho^5$,
and $A^6$ mesons. The `x's are the $\omega^1$, $f^2$, $\omega^3$,
$f^4$, and $f^6$. We have added one to the value of
the angular momentum $J$ in Figure~\ref{regge figure}
to account for the contribution of the spin of the quarks.

We have chosen a range of values for the quark masses
in Fig.~\ref{regge figure} in order to give
a qualitative picture of the dependence of the
Regge trajectory on the quark mass.
Since \eqnlessref{J result} does not include the contribution
of quark fluctuations to the Regge trajectory, this formula
is incomplete. We are now in the process of
including the quark degrees of freedom in the functional integral
\eqnlessref{param measure}. The boundary $\Gamma$ of the sheet
$\tilde x^\mu$ becomes dynamical, and couples to the string fluctuations.
It is clear that a calculation of the contribution of these
degrees of freedom is essential to understanding why the classical
formula for Regge trajectories works so well.

\section{Summary and Conclusions}

The primary results of the paper are \eqnlessref{param measure}
and \eqnlessref{J result}.
We have expressed the path integral $W[\Gamma]$
\eqnlessref{originalpartition} of a renormalizable
quantum field theory having classical vortex solutions
as the path integral formulation of an effective string
theory of vortices \eqnlessref{param measure}.
This theory describes the two transverse fluctuations
of the vortex at scales larger than the inverse
mass of the lightest particle in the field
theory. Our method is applicable to any
field theory containing vortex solutions.

Using the string representation of $W[\Gamma]$,
we carried out a semiclassical expansion of
the effective action $-i\log W[\Gamma]$ about
a classical solution of the effective string theory.
We calculated the contribution of these string
fluctuations explicitly for the case where the
worldline $\Gamma$ is generated by the
trajectory of a quark--antiquark pair
separated by a distance $R$. We are now calculating the
contribution to the effective action $-i\log W[\Gamma]$
due to the quantum fluctuations of the boundary.

\section*{Acknowledgments}

We would like to thank N. Brambilla for very helpful
conversations. This work was supported in part
by the U. S. Department of Energy grant
DE-FG03-96ER40956.

\appendix

\section{Notation and the Curvature of the Vortex}

\label{notation appendix}

We describe the string worldsheet by the function $\tilde x^\mu(\xi)$
of the coordinates $\xi$. The physics of the vortex should be
independent of the coordinate system we choose, so we require the
theory to be invariant under a reparameterization of the
coordinates, $\xi \to \tilde\xi(\xi)$. The tangent vectors
to the vortex worldsheet are defined by taking derivatives of
$\tilde x^\mu(\xi)$,
\begin{equation}
t^\mu_a(\xi) \equiv \partial_a \tilde x^\mu(\xi) \,,
\end{equation}
where $\partial_a = \partial/\partial\xi^a$ is a partial
derivative with respect to one of the vortex coordinates.
The induced metric on the worldsheet $\tilde x^\mu(\xi)$ is
\begin{equation}
g_{ab} \equiv t^\mu_a t_{\mu b} \,.
\label{metric appendix def}
\end{equation}
It is also convenient to define the square root of
the determinant of the metric,
\begin{equation}
\sqrt{-g} = \sqrt{-\frac{1}{2} \epsilon^{ab} \epsilon^{cd} g_{ac} g_{bd}} \,.
\end{equation}
We use the $t^\mu_a$ to define an antisymmetric
tensor which describes the orientation of the string
worldsheet,
\begin{equation}
t^{\mu\nu} \equiv \frac{\epsilon^{ab}}{\sqrt{-g}} t^\mu_a t^\nu_b \,.
\label{t mu nu def}
\end{equation}
This quantity was defined by Polyakov~\cite{morePolyakov}.
It is the projection of the antisymmetric tensor $\epsilon^{ab}$
into the space of four dimensional tensors. This tensor
defines the orientation of the two dimensional vortex worldsheet
in four space. The quantity \eqnlessref{t mu nu def} is also
independent of the coordinate parameterization of the
worldsheet $\tilde x^\mu(\xi)$.

We now describe the curvature of the vortex worldsheet.
We do this by taking covariant derivatives of the tangent vectors.
The covariant derivative of $t^\mu_a$ is
\begin{equation}
\nabla_b t^\mu_a = \partial_b t_\mu^a - \Gamma^c_{ab} t^\mu_c \,,
\end{equation}
where the $\Gamma^c_{ab}$ are Christoffel symbols,
\begin{equation}
\Gamma^c_{ab} = \frac{1}{2} g^{cd} \left( \partial_a g_{bd}
+ \partial_b g_{ad} - \partial_d g_{ab} \right) \,,
\label{Gamma def}
\end{equation}
The covariant derivatives of the tangent vectors are
orthogonal to the worldsheet,
\begin{equation}
t_{\mu c} \nabla_a t^\mu_b = t_{\mu c} \left( \partial_a t^\mu_b
- \Gamma^d_{ab} t^\mu_d \right) = 0 \,.
\label{no tangent covariant}
\end{equation}
This identity is derived using the definition
\eqnlessref{Gamma def} of the Christoffel symbols,
the definition \eqnlessref{metric appendix def}
of the metric, and the relationship between derivatives
of different $t^\mu_a$,
\begin{equation}
\partial_a t^\mu_b = \partial_b t^\mu_a = \partial_a \partial_b
\tilde x^\mu \,.
\label{tangent vector derivatives}
\end{equation}

The covariant derivatives of the tangent vectors are normal
to the string worldsheet. We therefore define a basis of
normal vectors $n^\mu_A$, which satisfy the conditions
\begin{equation}
n_{\mu A} t^\mu_a = 0 \,,\,\,\,\,\, n_{\mu A} n^\mu_B = \delta_{AB} \,.
\label{n def}
\end{equation}
The $n^\mu_A$ are an orthonormal basis for the vectors normal
to the worldsheet. \eqnref{no tangent covariant} implies that
\begin{equation}
\nabla_a t^\mu_b = n^\mu_A \scrK^A_{ab}
\label{curvature def}
\end{equation}
for some tensor $\scrK^A_{ab}$. The tensor $\scrK^A_{ab}$
is called the extrinsic curvature tensor of the string worldsheet.
With the definition \eqnlessref{curvature def}, the
curvature tensor $\scrK^A_{ab}$ is
\begin{equation}
\scrK^A_{ab} \equiv n_\mu^A \nabla_a t^\mu_b
= n_\mu^A \partial_a \partial_b x^\mu \,.
\end{equation}
It is symmetric in the indices $a$ and $b$ due to the
relationship \eqnlessref{tangent vector derivatives} between
derivatives of tangent vectors.

The extrinsic curvature of the string worldsheet can also
be described using derivatives of the normal vectors. The
orthogonality of the $t_{\mu a}$ and the $n^\mu_A$ implies
\begin{equation}
t_{\mu b} \partial_a n^\mu_A = - \scrK^A_{ab} \,.
\end{equation}
Therefore, the derivatives of the normal vectors can be written as
\begin{equation}
\partial_a n^\mu_A = - t^{\mu b} \scrK^A_{ab} + n^\mu_B \scrA^{AB}_a \,.
\end{equation}
The tensor $\scrA^{AB}_a$ is called the torsion, and it describes
the twisting of the basis of normal vectors as we move along
the worldsheet. The torsion depends on our choice of
the $n^\mu_A$, so we will choose them so that the torsion
is zero. This is done by requiring that the $n^\mu_A$
satisfy the differential equation
\begin{equation}
\partial_a n^\mu_A = n^\nu_A (\nabla_a t_{\nu b}) t^{\mu b} \,.
\label{torsion free}
\end{equation}
The equation \eqnlessref{torsion free} is equivalent to the
statement $\scrA^{AB}_b = 0$. It is consistent with the
conditions \eqnlessref{n def} which define the normal vectors.
As long as the normal vectors have an orthonormal basis at
one point, \eqnref{torsion free} guarantees they will be orthonormal
in a neighborhood of that point. Therefore, it is always
possible to find a local,
orthonormal, torsion free basis for the normal vectors.

There is one additional property of the normal vectors
we will use. The antisymmetric combination of the
normal vectors is (with proper ordering) equal to the
dual of the worldsheet orientation tensor $t^{\mu\nu}$
\eqnlessref{t mu nu def},
\begin{equation}
\epsilon^{AB} n^\mu_A n^\nu_B = \tilde t^{\mu\nu} \,,
\label{dual normal}
\end{equation}
where
\begin{equation}
\tilde t^{\mu\nu} = \frac{1}{2} \epsilon^{\mu\nu\alpha\beta}
t_{\alpha\beta} \,.
\end{equation}
The relationship \eqnlessref{dual normal} can be understood
by noting that any antisymmetric tensor is of the form
\begin{equation}
A^{\mu\nu}(\xi) = T(\xi) t^{\mu\nu} + N(\xi) \epsilon^{AB} n^\mu_A
n^\nu_B + M^{Aa} \left( n^\mu_A t^\nu_a - n^\nu_A t^\mu_a \right) \,.
\end{equation}
The tensor $\tilde t^{\mu\nu}$ is orthogonal to the $t^\mu_a$,
so it must be proportional to $\epsilon^{AB} n^\mu_A n^\nu_B$.
Squaring both of these tensors gives
\begin{equation}
\left( \tilde t^{\mu\nu} \right)^2 = \left( \epsilon^{AB} n^\mu_A n^\nu_B
\right)^2 = 2 \,.
\end{equation}
Therefore, $\tilde t^{\mu\nu}$ and $\epsilon^{AB} n^\mu_A n^\nu_B$
are equal up to an overall sign, which is fixed by choosing
an appropriate ordering for the normal vectors.

\section{Discussion of $J_\perp$}

\label{J_perp FP appendix}

The Jacobian $J_\perp$ \eqnlessref{J_perp expressed}
is a Faddeev--Popov determinant,
because fixing the position of the string
in the field integrals is analogous to fixing
a gauge in a gauge theory.
In the string action, we fix the degrees
of freedom which generate the transformation
\begin{equation}
\tilde x^\mu(\xi) \to \tilde x^\mu(\xi) + \delta x_\perp^A(\xi)
n_{\mu A}(\xi)
\end{equation}
which displaces the vortex.

The Jacobian $J_\perp$ is analogous to the Faddeev--Popov
determinant in a gauge theory.
In a gauge theory, where the $\delta$ function fixes the
symmetry generated by the transformation
\begin{equation}
A_\mu \to U A_\mu U^{-1} + U \partial_\mu U^{-1} \,,
\end{equation}
the Faddeev--Popov determinant appears as a normalization for
the $\delta$ function~\cite{Peskin+Schroder},
\begin{equation}
Z_{{\rm gauge}} = \int \scrD A^\mu \delta\left[F(A^\mu)\right] \Delta_{FP}
e^{-S} \,,
\label{gauge part}
\end{equation}
where
\begin{equation}
\Delta_{FP}^{-1} = \int \scrD U \delta\left[F(A^\mu)\right] \,.
\end{equation}
The Wilson loop \eqnlessref{gauge part} is
analogous to our equation \eqnlessref{Seff} for the
effective action. The determinant $\Delta_{FP}$ is
analogous to $J_\perp$.

In the gauge theory, the Faddeev--Popov method is used to remove
nonphysical degrees of freedom from the problem. The
$\delta$ function is inserted in the path integral
eqnlessref{gauge part} to fix the fields in
some particular gauge. This creates an integral over all
gauges which appears as a normalization factor, and is
removed. The $\delta$ function in \eqnref{Seff}, on the
other hand, fixes the position of the vortex sheet, which
is a physical degree of freedom.

\section{Evaluation of $S_1$}

\label{luscher appendix}

We want to evaluate the term,
\begin{equation}
S_1 = i \Tr\ln[-\nabla^2] \,,
\end{equation}
in the effective action for a general string
worldsheet. We work in coordinates
$r$ and $t$, such that $t$ is the time in the lab frame,
and $r$ is orthogonal to $t$ ($g_{rt} = 0$). In these
coordinates, the functional integral \eqnlessref{W_1 int} for
$e^{iS_1}$ takes the form
\begin{equation}
e^{iS_1} = \int \scrD f_1 \scrD f_2 \exp\left\{
-i\int dt \int_{-R_1}^{R_2} dr \sqrt{-g_{tt} g_{rr}}
\left[ g^{tt} \left(\frac{\partial f^i}{\partial t}\right)^2
+ g^{rr} \left(\frac{\partial f^i}{\partial r}\right)^2 \right]
\right\}\,.
\end{equation}

We consider the case where the metric is independent of $t$,
and we make the coordinate transformation $r \to x$ defined by
\begin{equation}
\frac{dx}{dr} = \sqrt{\frac{g_{rr}}{-g_{tt}}} \,, \kern 1 in
x \big|_{r=0} = 0 \,.
\end{equation}
The coordinate $x$ runs from $-X_1$ to $X_2$,
\begin{eqnarray}
X_1 &=& \int_{-R_1}^0 dr \sqrt{\frac{g_{rr}}{-g_{tt}}}
\nonumber \\
X_2 &=& \int_0^{R_2} dr \sqrt{\frac{g_{rr}}{-g_{tt}}} \,.
\end{eqnarray}
In these coordinates, the length of the
string is $X_1 + X_2 = R_p$, the proper length of the string,
\begin{equation}
R_p = \int_{-R_1}^{R_2} dr \sqrt{\frac{g_{rr}}{-g_{tt}}} \,.
\label{R_p appendix}
\end{equation}
In the coordinate system $(x,t)$, the metric is conformal,
\begin{eqnarray}
g_{xx} &=& \left(\frac{dx}{dr}\right)^{-2} g_{rr} = - g_{tt} \,,
\nonumber \\
g_{xt} &=& \left(\frac{dx}{dr}\right)^{-1} g_{rt} = 0 \,,
\end{eqnarray}
and
\begin{equation}
e^{iS_1} = \int \scrD f_1 \scrD f_2 \exp\left\{
\int dt \int_{-X_1}^{X_2}  dx \left[ -\left(\frac{\partial f^i}
{\partial t}\right)^2 + \left(\frac{\partial f^i}{\partial x}\right)^2
\right] \right\} \,.
\label{W_1 of x and t}
\end{equation}
We evaluate \eqnlessref{W_1 of x and t} in a manner analogous to
our treatment of $S_2$ in section~\ref{W_2 section}. We
Fourier transform in both space and time, introducing
variables $\nu$ and $k_n = \pi n/R_p$.
This transformation puts the action in \eqnlessref{W_1 of x and t}
in a diagonal form. Doing the $f_1$ and $f_2$ integrals gives
\begin{equation}
S_1 = \sum_{\nu,n}\ln\left[\nu^2 + \left(\frac{\pi n}{R_p}\right)^2
\right] \,,
\label{W_1 sum}
\end{equation}
where we have Wick rotated $\nu \to -i\nu$ to avoid the
poles at $\nu = \pm \pi n/R_p$.

The length $R_p$ is
just the classical string energy $E_{cl}$ divided
by the string tension $\sigma$, since $E_{cl}$ is
\begin{eqnarray}
E_{cl} &=& -\sigma \frac{1}{T} \int d^2\xi
\frac{\partial}{\partial \dot x^0} \sqrt{-g}
\nonumber \\
&=& \sigma \frac{1}{T} \int dt
\int_{-R_1}^{R_2} dr \sqrt{\frac{g_{rr}}{-g_{tt}}}
\nonumber \\
&=& \sigma R_p \,.
\label{E sigma}
\end{eqnarray}
The quantity $R_p = E_{cl}/\sigma$ is the length of
the string measured in local co-moving coordinates, which are
at rest with respect to the string. This is different from the
string length $R$ in the laboratory frame.

We will regulate $S_1$ using the results of LSW.
Their result for $S_{{\rm reg}}$ is the following,
\begin{eqnarray}
S_{{\rm reg}} &=& - \frac{d-2}{4\pi} A({\cal C})
\sum_j \epsilon_j {\cal M}_j^2 \ln {\cal M}_j^2
- \frac{d-2}{4} L({\cal C}) \sum_j \epsilon_j {\cal M}_j
\nonumber \\
& & + \left( \frac{d-2}{6} - \frac{1}{4\pi} \int d^2\xi \sqrt{-g}
{\cal R} \right) \sum_j \epsilon_j \ln {\cal M}_j^2 + S_{{\rm PV}} \,,
\label{Luscher reg}
\end{eqnarray}
where $d$ is the number of dimensions, $A({\cal C})$ is
the area of the string worldsheet, $L({\cal C})$ is the length
of its boundary, and ${\cal R}$ is the scalar curvature of the sheet.
The ${\cal M}_j$ are regulator masses, and the $\epsilon_j$ are
appropriate coefficients. The final term, $S_{{\rm PV}}$,
is finite in the limit where the ${\cal M}_j \to \infty$.

LSW evaluate the finite term $S_{{\rm PV}}$ only
for a straight string of length $R$
with fixed ends. In this case, the area of the sheet
$A({\cal C}) = RT$, the length of the boundary
$L({\cal C}) = 2T$, and the curvature of the sheet is
zero. They then obtained the explicit contribution
to the heavy quark potential:
\begin{equation}
- \lim_{T\to\infty} \frac{1}{T} S_{{\rm reg}}
= \frac{1}{2\pi} R \sum_j \epsilon_j {\cal M}_j^2 \ln {\cal M}_j^2
+ \sum_j \epsilon_j {\cal M}_j - \frac{\pi}{12 R} \,.
\label{luscher reg}
\end{equation}
The first term in \eqnlessref{luscher reg} renormalizes
the string tension. The second renormalizes the quark mass.
The third is the well known L\"uscher term in the heavy quark
potential.

Since the extrinsic curvature vanishes for a flat sheet,
we can identify the result \eqnlessref{luscher reg} with
our expression \eqnlessref{W_1 sum} for $S_1$, with
$R_p$ replaced by $R$:
\begin{equation}
- \lim_{T\to\infty} \frac{1}{T} \sum_{\nu,n}\ln\left[\nu^2
+ \left(\frac{\pi n}{R}\right)^2 \right]
= \frac{1}{2\pi} R \sum_j \epsilon_j {\cal M}_j^2 \ln {\cal M}_j^2
+ \sum_j \epsilon_j {\cal M}_j - \frac{\pi}{12 R} \,.
\label{flat eigenvalue sum}
\end{equation}
\eqnref{flat eigenvalue sum} tells us how to regulate $S_1$.
Replacing $R$ by $R_p$ in \eqnlessref{flat eigenvalue sum} gives
the regulated form of $S_1$:
\begin{equation}
-\lim_{T\to\infty} \frac{1}{T} S_1 =
\frac{1}{2\pi} R_p \sum_j \epsilon_j {\cal M}_j^2 \ln {\cal M}_j
+ \sum_j \epsilon_j {\cal M}_j - \frac{\pi}{12 R_p} \,.
\label{modified luscher}
\end{equation}

The first term in \eqnlessref{modified luscher}
is still a string tension renormalization, since both the
string tension contribution to the energy
\eqnlessref{E sigma} and the first term in
\eqnref{modified luscher} are proportional to $R_p$,
The second term in \eqnref{modified luscher} is, as before,
 a renormalization of the quark mass. The finite part of
the contribution of $S_1$ to the action is
\begin{equation}
S_1 \Big|_{{\rm finite part}} = T \frac{\pi}{12 R_p} \,.
\label{luscher R_p appendix}
\end{equation}
This is the result stated in section \ref{regularization section}.
The result \eqnlessref{luscher R_p appendix} is the L\"uscher
term, with the distance $R$ between the quarks replaced by
the proper length $R_p$ of the string. Our result
is \eqnlessref{R_p appendix}, the derivation of $R_p$.

\section{Cutoff Dependence of $S_2$}

\label{log cutoff appendix}

In this appendix, we show that
the divergent part of $S_2$ for a general sheet
is proportional to the integral of the scalar curvature
${\cal R}$. This agrees with the logarithmic divergence
\eqnlessref{Luscher reg} derived
by LSW by other means.  To obtain the divergent part of $S_2$,
we carry out a Fourier transform with respect to
the variable $x$. For functions defined on the
interval $-X_1 < x < X_2$,
the $\delta$ function can be expressed as a sum of sines,
\begin{equation}
\delta(x-x') = \frac{2}{R_p} \sum_{n=1}^\infty
\sin\left(k_n (x + X_1)\right)
\sin\left(k_n (x' + X_1)\right) \,,
\end{equation}
where $k_n = \pi n/R_p$. The Fourier transform of an operator
of the form $-\partial^2/\partial x^2 + U(x)$ can then be written
\begin{eqnarray}
\langle k_m | -\frac{\partial^2}{\partial x^2} + U(x) | k_n \rangle
&=& \frac{2}{R_p} \int_{-X_1}^{X_2} dx
\sin\left(k_m (x + X_1)\right)
\left( -\frac{\partial^2}{\partial x^2} + U(x) \right)
\sin\left(k_n (x + X_1)\right)
\nonumber \\
&=& k_n^2 \delta_{n,m} + \frac{2}{R_p} \int_{-X_1}^{X_2} dx
\sin\left(k_m (x + X_1)\right)
\sin\left(k_n (x + X_1)\right) U(x) \,.
\nonumber \\
\label{fourier transform}
\end{eqnarray}

Using the formula \eqnlessref{fourier transform} to
evaluate \eqnlessref{W_2 given} gives
\begin{eqnarray}
S_2 &=& \frac{T}{2} \Tr_n \Bigg[ \sqrt{k_n^2 \delta_{n,m} \delta_{AB}
- \frac{2}{R_p} \int_{-X_1}^{X_2} dx
\sin\left(k_m (x + X_1)\right)
\sin\left(k_n (x + X_1)\right)
\sqrt{-\bar g} \bar\scrK^A_{ab} \bar\scrK^{Bab}}
\nonumber \\
& & - k_n \delta_{n,m} \delta_{AB} \Bigg] \,,
\label{Fourier transformed}
\end{eqnarray}
The trace is over indices
$A,B$ which run from 1 to 2, and indices $n,m$ which run from
1 to $\infty$. The trace is cutoff at $k_n = M$,
the mass of the vector particle in the original field theory.

We expand \eqnlessref{Fourier transformed}
for large $k_n$ and obtain the cutoff dependent part of $S_2$,
\begin{equation}
S_2 = - \frac{T}{4} \sum_{n=1}^{M R_p/\pi} \frac{1}{R_p k_n}
\int_{-X_1}^{X_2} dx \sqrt{-\bar g} \left(\bar\scrK^A_{ab}\right)^2
+ \hbox{finite.}
\end{equation}
The term $\left(\bar\scrK^A_{ab}\right)^2$ is
equal to minus the scalar curvature ${\cal R}$,
\begin{equation}
{\cal R} = \left(\bar\scrK^{Aa}_a\right)^2
- \left(\bar\scrK^A_{ab}\right)^2 \,,
\end{equation}
since the equation of motion \eqnlessref{class string simple}
implies $\bar\scrK^{Aa}_a = 0$. The cutoff dependent part of
$S_2$ is therefore
\begin{equation}
S_2 = \frac{T}{4} \sum_{n=1}^{M R_p/\pi} \frac{1}{\pi n}
\int_{-X_1}^{X_2} dx \sqrt{-\bar g} {\cal R} + \hbox{finite.}
\label{W_2 divergence}
\end{equation}
\eqnref{W_2 divergence} agrees with the result of LSW
for the leading semiclassical logarithmic divergence.

\section{Evaluation of $S_2$}

\label{eigenfunctions and sum}

We want to evaluate $S_2$,
\begin{equation}
S_2 = \frac{T}{2} \left( \Tr\sqrt{-\frac{\partial^2}{\partial x^2}
\delta_{AB} - \sqrt{-\bar g} \scrK^A_{ab} \scrK^{Bab}}
- \Tr\sqrt{-\frac{\partial^2}{\partial x^2} \delta_{AB}} \right)
\label{V trace}
\end{equation}
for the fluctuations about a straight string of length
$R$ rotating with angular velocity $\omega$.
To evaluate this, we must determine
the value of the extrinsic curvature $\scrK^A_{ab}$. The
definition of the extrinsic curvature is
\begin{equation}
\scrK^A_{ab} = n_\mu^A \partial_a \partial_b x^\mu \,.
\end{equation}
The string $x^\mu$ is
\begin{equation}
x^\mu(x,t) = t \ehat_0^\mu + \frac{1}{\omega} \sin(\omega x)
\left( \cos(\omega t) \ehat_1^\mu + \sin(\omega t) \ehat_2^\mu \right) \,.
\end{equation}
The $\ehat_i^\mu$ are a basis of orthonormal
unit vectors in Minkowski space.
The $n^\mu_A$ are a basis for the vectors normal to
$x^\mu$. We choose the basis
\begin{eqnarray}
n^\mu_1 &=& \ehat_3^\mu
\nonumber \\
n^\mu_2 &=& \tan(\omega x) \ehat_0^\mu
+ \sec(\omega x) \left(-\sin(\omega t) \ehat_1^\mu
+ \cos(\omega t) \ehat_2^\mu \right) \,.
\end{eqnarray}
With this choice for the $n^\mu_A$, $\scrK^1_{ab}$ is zero,
because the $\ehat_3^\mu$ component of $x^\mu$ is zero.
The only nonzero component of $\scrK^A_{ab} \scrK^{Bab}$ is
\begin{equation}
\sqrt{-\bar g} \left(\scrK^2_{ab}\right)^2 = - 2 \omega^2 \sec^2 \omega x \,.
\label{nonzero k}
\end{equation}

Now that we know what $\scrK^A_{ab}$ is, we can
evaluate \eqnlessref{V trace}.
Inserting \eqnlessref{nonzero k} into \eqnlessref{V trace} gives
\begin{equation}
S_2 = \frac{T}{2} \left( \Tr\sqrt{-\frac{\partial^2}{\partial x^2}
+ 2 \omega^2 \sec^2 \omega x} - \Tr\sqrt{-\frac{\partial^2}{\partial x^2}}
\right) \,.
\label{final rotating potential}
\end{equation}
The traces in \eqnlessref{final rotating potential} are defined
as sums over the eigenvalues of the given operators.
Replacing the traces with explicit sums gives
\begin{equation}
S_2 = \frac{T}{2} \sum_{n=1}^{\frac{\Lambda R_p}{\pi}}
\left( \sqrt{\lambda_n} - \frac{\pi n}{R_p} \right) \,,
\label{sum to do}
\end{equation}
where
\begin{equation}
R_p = \frac{\arcsin v}{v} R \,.
\end{equation}
The eigenvalues $\lambda_n$ are determined by the eigenfunction
equation
\begin{equation}
\left( -\frac{\partial^2}{\partial x^2}
+ 2 \omega^2 \sec^2 \omega x \right) \psi_n(x) = \lambda_n \psi_n(x) \,,
\label{eigenfunction equation}
\end{equation}
with the boundary conditions $\psi_n(\pm R_p/2) = 0$.
The difference between the traces in \eqnref{final rotating potential}
is logarithmically dependent on the cutoff $\Lambda$ (the mass of the
dual gluon).

\eqnref{eigenfunction equation} has the form of the
Schr\"oedinger equation, with the potential $2\omega^2 \sec^2\omega x$.
This potential is an analytic continuation of the
potential $2\omega^2\sech^2\omega x$, whose eigenfunctions
can be expressed terms of hypergeometric functions
\cite{Landau+Lifshitz}. Using this result, we find the
eigenfunctions
\begin{equation}
\psi_n(x) =
  \begin{cases}
    \sqrt{\lambda_n} \cos\left(\sqrt{\lambda_n} x\right)
    + \omega \tan\omega x \sin\left(\sqrt{\lambda_n} x\right) &
    \text{for n odd,} \\
    \sqrt{\lambda_n} \sin\left(\sqrt{\lambda_n} x\right)
    - \omega \tan\omega x \cos\left(\sqrt{\lambda_n} x\right) &
    \text{for n even.}
  \end{cases}
\label{eigenfunctions}
\end{equation}
The eigenvalues $\lambda_n$ are
\begin{equation}
\lambda_n = \left(\frac{\left(\pi n + 2 \alpha_n\right) v}
{R \arcsin v}\right)^2 \,,
\end{equation}
where $\alpha_n$ satisfies the transcendental equation
\begin{equation}
\frac{\frac{\pi}{2} n + \alpha_n}{\arcsin v}
= \frac{v}{\sqrt{1 - v^2}} \cot\alpha_n \,,
\label{alpha def}
\end{equation}
and $0 < \alpha_n < \pi/2$. There is no $n=0$ eigenvalue,
despite the fact that $\alpha_0 = \arcsin v$
satisfies \eqnlessref{alpha def}, because the corresponding
eigenvalue $\lambda_0 = \omega^2$ makes $\psi_n$
zero everywhere.

We will carry out the sum \eqnlessref{sum to do},
\begin{equation}
S_2 = \frac{T}{2} \sum_{n=1}^{\frac{\Lambda R_p}{\pi}}
\left( \sqrt{\lambda_n} - \frac{\pi n}{R_p} \right) \,,
\label{W_2 sum before contour}
\end{equation}
by converting it to a contour integral. We will find
a function $F_\lambda(z)$ which has zeros whenever
$z = \pm \sqrt{\lambda_n}$. We will find another function
$F_{R_p}(z)$ which has zeros whenever $z = \pm \pi n/R_p$.
We will then define a function $F_{int}(z)$,
\begin{equation}
F_{int}(z) = \frac{d\ln F_\lambda(z)}{dz}
- \frac{d\ln F_{R_p}(z)}{dz} \,.
\end{equation}
The function $F_{int}(z)$ has poles of residue $1$ when
$z = \pm\sqrt{\lambda_n}$ and poles of residue $-1$ when
$z = \pm \pi n/R_p$. We then rewrite the sum
\eqnlessref{W_2 sum before contour} as a contour
integral,
\begin{equation}
S_2 = \frac{T}{4\pi i} \int dz z F_{int}(z) \,.
\label{W_2 contour abstract}
\end{equation}
The contour of the integral \eqnlessref{W_2 contour abstract}
lies along the imaginary axis, and on a semicircle at $|z| = \Lambda$
with the real part of $z$ positive.

To write $S_2$ as the contour integral \eqnlessref{W_2 contour
abstract}, we need to find the functions
$F_\lambda(z)$ and $F_{R_p}(z)$. The function $F_{R_p}(z)$ is
\begin{equation}
F_{R_p}(z) = \sin(R_p z) \,,
\end{equation}
which is zero for $z = \pm \pi n/R_p$. We find
$F_\lambda(z)$ by recalling that
the eigenfunctions \eqnlessref{eigenfunctions} vanish
at $x = \pm R_p/2$. Therefore,
\begin{equation}
F_{{\rm odd}}(z) = z \cos\left(\frac{R_p}{2} z\right)
+ \omega \tan\left(\frac{R_p}{2} \omega\right)
\sin\left(\frac{R_p}{2} z\right) \,,
\end{equation}
has zeros at $z = \sqrt{\lambda_n}$ for $n$ odd, and
\begin{equation}
F_{{\rm even}}(z) = z \sin\left(\frac{R_p}{2} z\right)
- \omega \tan\left(\frac{R_p}{2} \omega\right)
\cos\left(\frac{R_p}{2} z\right) \,,
\end{equation}
has zeros at $z = \sqrt{\lambda_n}$ for $n$ even. Thus,
\begin{eqnarray}
F_\lambda(z) &=& \frac{F_{{\rm odd}}(z) F_{{\rm even}}(z)}{z^2 - \omega^2}
\nonumber \\
&=& \frac{1}{2} \frac{z^2 \sin(R_p z) - 2\omega z \tan\left(
\frac{R_p}{2} \omega\right) \cos(R_p z)
- \omega^2 \tan^2\left(\frac{R_p}{2} \omega\right) \sin(R_p z)}
{z^2 - \omega^2} \,.
\end{eqnarray}
The factor $(z^2-\omega^2)^{-1}$ removes nonphysical
zeros which appear because $F_{{\rm even}}(\pm\omega) = 0$.
These zeros correspond to the $n=0$ ``eigenfunction'' which is zero
everywhere for $\lambda_0 = \omega^2$. The function
$F_{int}(z)$ is
\begin{equation}
F_{int}(z) = \frac{d}{dz} \ln\left[ \frac{z^2
- 2\omega z \tan\left(\frac{R_p}{2} \omega\right) \cot(R_p z)
- \omega^2 \tan^2\left(\frac{R_p}{2} \omega\right)}
{z^2 - \omega^2} \right] \,.
\label{F int found}
\end{equation}

Inserting \eqnlessref{F int found} in
\eqnlessref{W_2 contour abstract} and integrating by parts gives
\begin{equation}
S_2 = - \frac{T}{4\pi i} \int dz \ln\left[\frac{z^2 - 2\omega z
\tan\left(\frac{R_p}{2} \omega\right) \cot\left(R_p z\right)
- \omega^2 \tan^2\left(\frac{R_p}{2} \omega\right)}{z^2-\omega^2}\right] \,.
\label{contour defined}
\end{equation}
Now, instead of having poles at $z = \sqrt{\lambda_n}$ and $z = \pi n/R_p$,
the integrand has branch points at these points. The branch cuts run
from $\sqrt{\lambda_n}$ to $\pi n/R_p$ for all $n$ along the real axis.
There is one branch cut for each value of $n$. Since the contour
either includes both $\sqrt{\lambda_n}$ and $\pi n/R_p$ or
excludes both these points, none of these branch cuts cross
the contour of integration.
There are no branch points at $z = \pm\omega$, since both the
numerator and denominator vanish there.
None of the branch cuts crosses the contour, so the contour is
still closed, and the integration by parts does not produce
a boundary term.

The contour of the integral \eqnlessref{contour defined}
lies on the imaginary axis and a semicircle passing through positive
real infinity. We rewrite \eqnlessref{contour defined}
as two integrals over the different pieces of
the contour. For large values of the cutoff $\Lambda$,
the action $S_2$ is
\begin{eqnarray}
S_2 &=& -\frac{T}{4\pi} \int_{-\Lambda}^{\Lambda}
dy \ln\left[\frac{y^2 + 2\omega y 
\tan\left(\frac{R_p}{2} \omega\right) \coth\left(R_p y\right)
+ \omega^2 \tan^2\left(\frac{R_p}{2} \omega\right)}{y^2+\omega^2}\right]
\nonumber \\
& & - T \frac{1}{4\pi} \int_{-\pi/2}^{\pi/2} d\theta \Lambda e^{i\theta}
\left[-2\frac{\omega}{\Lambda} e^{-i\theta}
\tan\left(\frac{R_p}{2} \omega\right) \cot\left(R_p \Lambda e^{i\theta}\right)
+ O(\Lambda^{-2}) \right] \,.
\end{eqnarray}
For large $\Lambda$, $\cot\left(R_p \Lambda e^{i\theta}\right)$
is proportional to the sign of $\theta$, so the $\theta$ integral vanishes.
The $y$ integral is symmetric under $y\to -y$. Changing variables
to $s = (y/\omega) \cot\left(\omega R_p/2\right)$ gives
\begin{equation}
S_2 = - T \frac{\omega}{2\pi} \tan\left(\frac{R_p}{2} \omega\right)
\int_0^{\frac{\Lambda}{\omega} \cot\left(\omega R_p/2\right)}
ds \ln\left[ \frac{s^2 + 2s \coth\left(R_p \omega
\tan\left(\frac{R_p}{2} \omega\right) s\right) + 1}
{s^2 + \cot^2\left(\frac{R_p}{2} \omega\right)}\right] \,.
\label{s integral}
\end{equation}

The numerator of in \eqnlessref{s integral} is approximately
$(s+1)^2$ for large values of $s$. We use this fact to extract
the divergent part of \eqnlessref{s integral}, getting
\begin{equation}
S_2 = - T \frac{2 v}{\pi R} \left[ v \gamma
\ln\left(\frac{\Lambda R}{2v^2 \gamma}\right)
+ v\gamma - \frac{\pi}{2} \right]
- T \frac{v^2 \gamma}{\pi R} f\left( v \right) \,.
\label{S_2 Lambda cutoff}
\end{equation}
We have replaced $R_p$ with its definition
\begin{equation}
R_p = \frac{2}{\omega} \arcsin v \,.
\end{equation}
$\gamma = (1-v^2)^{-1/2}$ is the quark boost factor.
The function $f(v)$ contains the rest of the integral,
\begin{equation}
f(v) \equiv \int_0^\infty ds \ln\left[\frac{s^2
+ 2 s \coth\left(2s v \gamma \arcsin v \right) + 1}
{(s+1)^2}\right] \,.
\end{equation}
For $v \to 1$, the asymptotic value of $f(v)$ is
\begin{equation}
f(v) \simeq \frac{1}{6 \gamma^2} \,.
\end{equation}

The cutoff $\Lambda$ used in \eqnlessref{S_2 Lambda cutoff}
is the cutoff in the $x$ coordinate. We must express $\Lambda$
in terms of the cutoff $M$ for the $r$ coordinate, which measures
physical distance. The cutoffs $\Lambda$ and $M$ are
related by the equation $\Lambda \delta x = M \delta r$, or
\begin{equation}
\Lambda = M \frac{dr}{dx} = M \gamma^{-1} \,.
\label{Lambda of M}
\end{equation}
Inserting \eqnlessref{Lambda of M} into \eqnlessref{S_2 Lambda cutoff}
gives
\begin{equation}
S_2 = - T \frac{2 v}{\pi R} \left[ v \gamma
\ln\left(\frac{M R}{2(\gamma^2 - 1)}\right)
+ v\gamma - \frac{\pi}{2} \right]
- T \frac{v^2 \gamma}{\pi R} f\left( v \right) \,.
\end{equation}
Using the classical equation of motion \eqnlessref{boundary condition}
then gives
\begin{equation}
S_2 = - T \frac{2 v}{\pi R} \left[ v \gamma
\ln\left(\frac{M m}{\sigma}\right)
+ v\gamma - \frac{\pi}{2} \right]
- T \frac{v^2 \gamma}{\pi R} f\left( v \right) \,.
\label{S_2 final}
\end{equation}

\end{document}